\documentclass[twocolumn]{aastex63} %linenumbers

\usepackage{float}
\usepackage{CJKutf8}
\usepackage{xcolor}
\usepackage{multirow}
\usepackage{graphicx}
\usepackage{amsmath}

\received{3 Mar 2021}
\revised{29 Aug 2021}
\accepted{29 Sept 2021}
\shorttitle{Constraining the IMF with Wolf–Rayet spectra}
\shortauthors{Liang et al.}

\graphicspath{{./}{figures/}}

\begin{document}

\title{Wolf–Rayet galaxies in SDSS-IV MaNGA. II.
  Metallicity dependence of the high-mass slope of the stellar initial mass function}

%\correspondingauthor{Cheng Li}
\email{Contact e-mail: ericfuhengliang@gmail.com (FHL); cli2015@tsinghua.edu.cn (CL)}

% This Chinese character package can be compiled with PDFLatex
\author[0000-0003-2496-1247]{
\begin{CJK*}{UTF8}{gbsn}
Fu-Heng Liang (梁赋珩)\end{CJK*}
}
\affiliation{Department of Astronomy, Tsinghua University, Beijing 100084, China}
\affiliation{Sub-department of Astrophysics, Department of Physics, University of Oxford, Keble Road, Oxford OX1 3RH, UK}

\author[0000-0002-8711-8970]{Cheng Li}
\affiliation{Department of Astronomy, Tsinghua University, Beijing 100084, China}

\author{Niu Li}
\affiliation{Department of Astronomy, Tsinghua University, Beijing 100084, China}

\author{Shuang Zhou}
\affiliation{Department of Astronomy, Tsinghua University, Beijing 100084, China}

\author{Renbin Yan}
\affiliation{Department of Physics \& Astronomy, University of Kentucky, Lexington, KY 40506, USA}

\author{Houjun Mo}
\affiliation{Department of Astronomy, University of Massachusetts Amherst, MA 01003, USA}

\author{Wei Zhang}
%\affiliation{National Astronomical Observatories, Chinese Academy of Sciences, 20A Datun Road, Chaoyang District, Beijing 100012, China}
\affiliation{CAS Key Laboratory of Optical Astronomy, National Astronomical Observatories, Chinese Academy of Sciences, Beijing 100101, China}

%\author{Others}
%\affiliation{Other institutions}

\begin{abstract}
  As hosts of living high-mass stars, Wolf–Rayet (WR) regions or WR
  galaxies are ideal objects for constraining the high-mass end of the
  stellar initial mass function (IMF).  We construct a large sample of
  910 WR galaxies/regions that cover a wide range of stellar
  metallicity (from Z$\sim$0.001 up to Z$\sim$0.03), by combining three catalogs of WR galaxies/regions 
  previously selected from the SDSS and SDSS-IV/MaNGA surveys. We
  measure the equivalent widths of the WR blue bump at
  $\sim4650$ {\AA} for each spectrum. 
  They are compared with
  predictions from stellar evolutionary models {\tt Starburst99} and {\tt BPASS}, with different IMF assumptions (high-mass slope $\alpha$ of the IMF ranging from 1.0 up to 3.3).
Both singular evolution and binary evolution are considered.
  We also use a Bayesian inference code to perform full spectral
  fitting to WR spectra with stellar population spectra %  the singular and binary
  from {\tt BPASS} as fitting templates. 
  We then make model selection among different $\alpha$ assumptions based on Bayesian evidence. These analyses have consistently led to a positive correlation of IMF high-mass slope $\alpha$ with stellar metallicity $Z$, i.e. with steeper IMF  (more bottom-heavy) at higher metallicities. Specifically, an IMF with $\alpha$=1.00 is preferred at the lowest metallicity (Z$\sim$0.001), and a Salpeter or even steeper IMF is preferred at the highest metallicity (Z$\sim$0.03). These conclusions hold even when binary population models are adopted.
\end{abstract}


\section{Introduction}

The stellar initial mass function (IMF) describes the mass
distribution of stars at birth. The IMF is of critical importance in
studies of both star formation and galaxy evolution, as it influences
most observable properties of stellar populations and galaxies.  As
originally proposed by \citet{Salpeter-55}, the IMF may be
characterized by a simple power-law function: 
%{\color{blue}
%%\mathrm{d}N/\mathrm{d}m = m^{-\alpha}
%\begin{equation*}
\[  \frac{\text{d}N}{\text{d}m}  { { \propto }} \ m^{-\alpha}, \]
%\end{equation*}
%}
where $m$ is the mass of a star, $N$ is the number of
stars within the mass range $m+\mathrm{d}m$, and $\alpha$ is the slope.  Numerous observational
constraints from later studies suggested that, the Salpeter power law form
holds only at high masses above a few ${\mathrm{M_\odot}}$, while the IMF at
lower masses can be described by a log-normal distribution
\citep{Chabrier-03}, or similarly by a double power-law form
\citep{Kroupa-01}. Whether the IMF is universal or sensitive to
environmental conditions of star formation has long been debated.
\citet{2010ARA&A..48..339B} 
showed an eight-parameter IMF containing low- and high-mass limits, low- and high-mass slopes, low- and high-mass breaks, and mean and dispersion of the log-normal distribution in the intermediate mass range.
They also 
reviewed early reports
of IMF variation and concluded that the IMF follows a universal
Salpeter index at the high-mass end ($\alpha=2.35$) and a shallower
slope at low masses with a range of index ($\alpha\sim1-\sim1.25$).

The universal slope at the high-mass end has been disputed in the past
decade, however, mainly based on studies of old stellar populations in
early-type galaxies or bulges of late-type galaxies. A variety of
techniques have been used, such as absorpition line spectroscopy
\citep[e.g.][]{2012ApJ...760...71C, 2015MNRAS.447.1033M, Conroy-vanDokkum-Villaume-17},
kinematic analysis \citep[e.g.][]{Cappellari-12,2017ApJ...838...77L},
and stellar population synthesis
\citep[e.g.][]{2018MNRAS.477.3954P,2019MNRAS.485.5256Z}.  These
studies have well established that the IMF slope at the massive end
varies with the central stellar velocity dispersion ($\sigma_\ast$),
with steeper IMF's at higher $\sigma_\ast$. More recently, using
integral field spectroscopy (IFS) from the Mapping
  Nearby Galaxies at Apache Point Observatory (MaNGA) survey \citep{2015ApJ...798....7B},
\citet{2018MNRAS.477.3954P} and \citet{2019MNRAS.485.5256Z} found the
IMF slope to be also correlated with stellar metallicity ($Z_\ast$, simplified as $Z$ in this paper) in
early-type galaxies. \citet{2019MNRAS.485.5256Z} further demonstrated
that $Z$ is a more fundamental parameter than $\sigma_\ast$ in
driving the variation of the high-mass slope of IMF,  which confirms earlier results of \citet{2015ApJ...806L..31M}. With MUSE IFS data, \citet{2019A&A...626A.124M} further proposes a link between the IMF and distribution of warm orbits.

In fact, the metallicity dependence of the IMF high-mass slope has
also been found in late-type galaxies, by analyzing spectra of
Wolf-Rayet (WR) galaxies \citep{Zhang-07}. WR 
% removed Guseva-Izotov-Thuan-00 in this line but keep the following mention. Reason is though they have similar plots as Zhang's, they actually explained them differently, not variation of the IMF.
galaxies are identified by significant  stellar emission lines from WR
stars, which are a rare population  of living massive stars at the
post-main-sequence stage,  believed to evolve from O-type stars with
an initial mass exceeding  $25\ \mathrm{M_\odot}$ \citep[at solar metallicity;][]{Crowther-07}. The WR
emission features in  integrated spectra of WR galaxies/regions should
in priciple  provide stringent constraints on the high-mass slope of
the IMF. {Moreover, the WR stellar population represent the youngest population with age less than 10 Myr. The constraints from WR population can provide important supplement to aforementioned studies on old stellar population, i.e. early-type galaxies or bulges of late-type galaxies.}
By analyzing  long-slit spectra of 39 Wolf-Rayet galaxies,
\citet{Guseva-Izotov-Thuan-00} found the WR emission features at the lowest metallicity bin cannot
be explained by the stellar population models of
\citet{Schaerer-Vacca-98} that assumed a Salpeter IMF, and a very
shallow IMF slope of $\alpha=1$ provided better matching between the model
and the data, though they did not attribute this phenomenon to variation of the IMF slope. Based on a
large sample of 174 WR galaxies identified from the Sloan Digital Sky
Survey \citep[SDSS;][]{York2000}, \citet{Zhang-07} extended their work by comparing the observed WR
spectral features with model predictions of \citet{Schaerer-Vacca-98}
for a wider range of IMF slopes. This analysis
led to a positive correlation of the IMF slope with metallicity, in
good agreement with other studies of early-type galaxies. 

This paper is the second of a series studies on the WR spectra of
galaxies in MaNGA. Following \citet[][hereafter Z07]{Zhang-07}, we attempt to constrain
the slope of the high-mass end of the IMF by comparing the spectroscopic
features of WR galaxies of different metallicities, with predictions
of stellar population models covering a range of IMF slope index.  We
concentrate on examining the metallicity dependence of the IMF slope,
but have significantly extended the work of Z07 in the following
aspects.  First, we have constructed a much larger sample of WR
spectra, including 910 WR regions from three existing large catalogs (Z07;
\citealt{Brinchmann-Kunth-Durret-08}, hereafter B08; \citealt{liang_2020}, the first paper in this series, hereafter Paper I). In
particular, the catalog of Paper I is based on the MaNGA
data, thus including WR regions not only at galactic centers but also
in the outer disk of galaxies. Second, we consider two different stellar
population models: the traditional models of \citet{Schaerer-Vacca-98}
in the {\tt Starburst} code, which allows comparisons with previous studies,
and the recently-developed models in the Binary Population and Spectral Synthesis ({\tt BPASS}) code
\citep{2017PASA...34...58E, 2018MNRAS.479...75S}. Particularly,
the {\tt BPASS} includes binary stellar populations, thus allowing
us to examine the effect of binary stars on the variation of IMF.
Finally, we perform full spectral fitting to all the WR spectra
in our sample, using the Bayesian inference code {\tt BIGS} developed
recently by \citet{2019MNRAS.485.5256Z}. The Bayesian evidence and
the model parameters inferred from the spectra provide a reliable way
to do model selection between models of different IMF slopes
and between models of singular and binary populations.

Besides the high-mass slope, other parameters of the IMF are also widely studied. Since our approach with extragalactic WR stellar population is only sensitive and accurate regarding the high-mass slope of the IMF, this paper is only focused on this. We point readers to more comprehensive reviews of the IMF parameters in \citet{2010ARA&A..48..339B} and \citet{2014prpl.conf...53O}, and also note that potential degeneracy between high-mass slope and high-mass break could exist \citep{2019ApJ...870...44H}.

The paper is arranged as follows. In \autoref{sec:data}, we describe
the combined WR catalog, the procedure of measuring the WR blue bump, the
stellar evolutionary models of {\tt Starburst99} and {\tt BPASS}, and
the procedure of applying {\tt BIGS} to WR spectra.  In
\autoref{sec:results}, we present the comparison of the observed WR
feautres with those in models. We discuss the results among literature in \autoref{sec:dis} and conclude in \autoref{sec:con}.

\section{Data and models}
\label{sec:data}

\subsection{The Wolf-Rayet catalog}
\label{sec:catalog}

We use three catalogs of WR galaxies for this work. 
The first one was constructed in Paper I using  spatially 
resolved spectroscopy from MaNGA. The other two catalogs were 
selected from SDSS, by Z07 and B08 independently, thus 
limited to the central $3^{\prime\prime}$ region of the 
galaxies. All the WR galaxies or regions are identified by the broad
emission feature (i.e. the blue WR bump) at around 
4650\AA, but in practice different authors have 
adopted slightly different procedure/criteria. We briefly 
describe the construction process of the catalogs, and 
refer the reader to the relevant papers for details
\citep{Zhang-07, Brinchmann-Kunth-Durret-08, liang_2020}.

The WR catalog of Z07 was constructed from the SDSS/Data Release (DR) 3 galaxy 
sample in a two-step method. Out of the $\sim3.7\times10^5$
galaxies from SDSS/DR3, star-forming galaxies with significant 
H$\mathrm{\epsilon}$ emission line were first selected as candidates, 
and then WR galaxies were selected by visually examining the 
spectrum of the candidates. The star-forming galaxies are 
classified on the diagram introduced by \citet{Baldwin-Phillips-Terlevich-81}, 
and the H$\mathrm{\epsilon}$ line is required to have 
an equivalent width $EW\mathrm{(H\epsilon)}>5$\AA. 
As pointed out by the authors, by selection this catalog 
must be biased to high-excitation H{\sc ii} regions and may have 
missed those WR galaxies with spectra of low signal-to-noise ratio ($S/N$). 
The Z07 catalog includes 174 WR galaxies. 

The WR catalog of B08 was based on a later SDSS sample, 
the SDSS/DR6, which includes spectra for nearly $8\times10^{5}$ 
galaxies. The selection of WR galaxies started with the subset of 
$\sim3\times10^5$ spectra with the equivalent width of 
the H${\mathrm{\beta}}$ emission line $>2$\AA. To identify 
candidate WR galaxies, the excess flux above the best-fit 
continuum in spectral regions around the WR features 
was calculated for each spectrum. All the spectra were 
then ordered by decreasing the excess flux, and the first 
$11,241$ spectra were visually examined. A sample of 570 
galaxies were finally selected as WR galaxies, of which 
101 are in common with the Z07 catalog. We have performed full 
spectral fitting to all the spectra in SDSS (see below) and 
found the 73 galaxies to present a similar distribution of $EW_{4650}$ to that of all WR spectra in our catalog. 
Therefore, we keep the 73 spectra in our catalog. 
% removed: The rest 73 galaxies from Z07 but not included in B08 were classified as possible WR galaxies by B08.
% actually 36/73 are possible WR detection in B08 and further 37/73 are null detection.

The WR catalog of Paper I  was constructed using the MaNGA 
MaNGA Product Launch (MPL)-7 sample, which contains 4688 datacubes for 4621 
unique galaxies and was released as a part of the fifteeth 
data release of SDSS \citep[DR15;][]{2019ApJS..240...23A}. 
A two-step searching scheme was adopted. First, for each 
galaxy, H{\sc ii} regions were identified using the two-dimensional 
map of extinction-corrected H$\mathrm{\alpha}$ surface brightness, and the 
spectra falling in each region were stacked to generate an
average spectrum with high spectral $S/N$. Next, for each region,  
the stellar component (the continuum plus absorption lines)
was derived by performing full spectral fitting to the stacked 
spectrum, and the starlight-subtracted spectrum was visually 
inspected. An H{\sc ii} region was identified to be a WR 
region if it presents a significant WR bump around 4650\AA, 
and a galaxy was identified as a WR galaxy if it contains 
at least one WR region. This procedure results in a sample of 
90 WR galaxies containing a total of 267 WR regions. 
The reader is referred to Paper I for detailed description 
of the selection process, as well as analyses of the global 
properties of the WR galaxies. For details about the MaNGA 
survey, the reader is referred to 
\citet{2017AJ....154...28B} for an overview of the SDSS-IV project, 
\citet{2006AJ....131.2332G} and \citet{2013AJ....146...32S} 
for the Sloan telescope and the BOSS spectrograph
with which the MaNGA data are obtained, 
\citet{2015ApJ...798....7B} for an overview of the MaNGA survey, 
\citet{2015AJ....149...77D} for MaNGA instrumentation, 
\citet{Wake-17} for MaNGA sample design, 
\citet{2015AJ....150...19L} for observing strategy, 
\citet{2016AJ....151....8Y} for flux calibration, 
\citet{2016AJ....152...83L} for data reduction pipeline, and 
\citet{2019ApJS..240...23A} for the SDSS DR15. 

Combining the three WR catalogs, we have a total of 910 unique 
WR spectra. As mentioned above, 101 SDSS-based WR spectra 
are commonly included in Z07 and B08, and we have removed the 
duplicate spectra before combining the two catalogs. For the MaNGA-based catalog, 
the majority of the WR galaxies that contain a WR region at 
their center were missed by either Z07 or B08, and only 7 
of such galaxies were selected into those earlier catalogs. 
As pointed out in Paper I, the relatively low spectral $S/N$ of the 
spectra was the main reason for the SDSS to have missed 
those WR galaxies. For each of the 7 common galaxies, we keep
both the stacked spectra of their WR regions from MaNGA and 
the single-fiber spectrum from SDSS, as the two spectra cover 
slightly different radii and have different spectral $S/N$. 
In addition, we find five galaxies were duplicated in the B08 
catalog due to repeated observations of some SDSS plates. 
Given the small number, the 
duplication in our catalog is expected to have little effect 
on our results. 
% removed: We keep the duplicated spectra of the five galaxies in our  catalog, considering that the repeated spectra may provide  consistency check for the data. 
% because we do not do/mention the consistency check at all in the following texts.

\begin{figure}
    \includegraphics[width=\columnwidth]{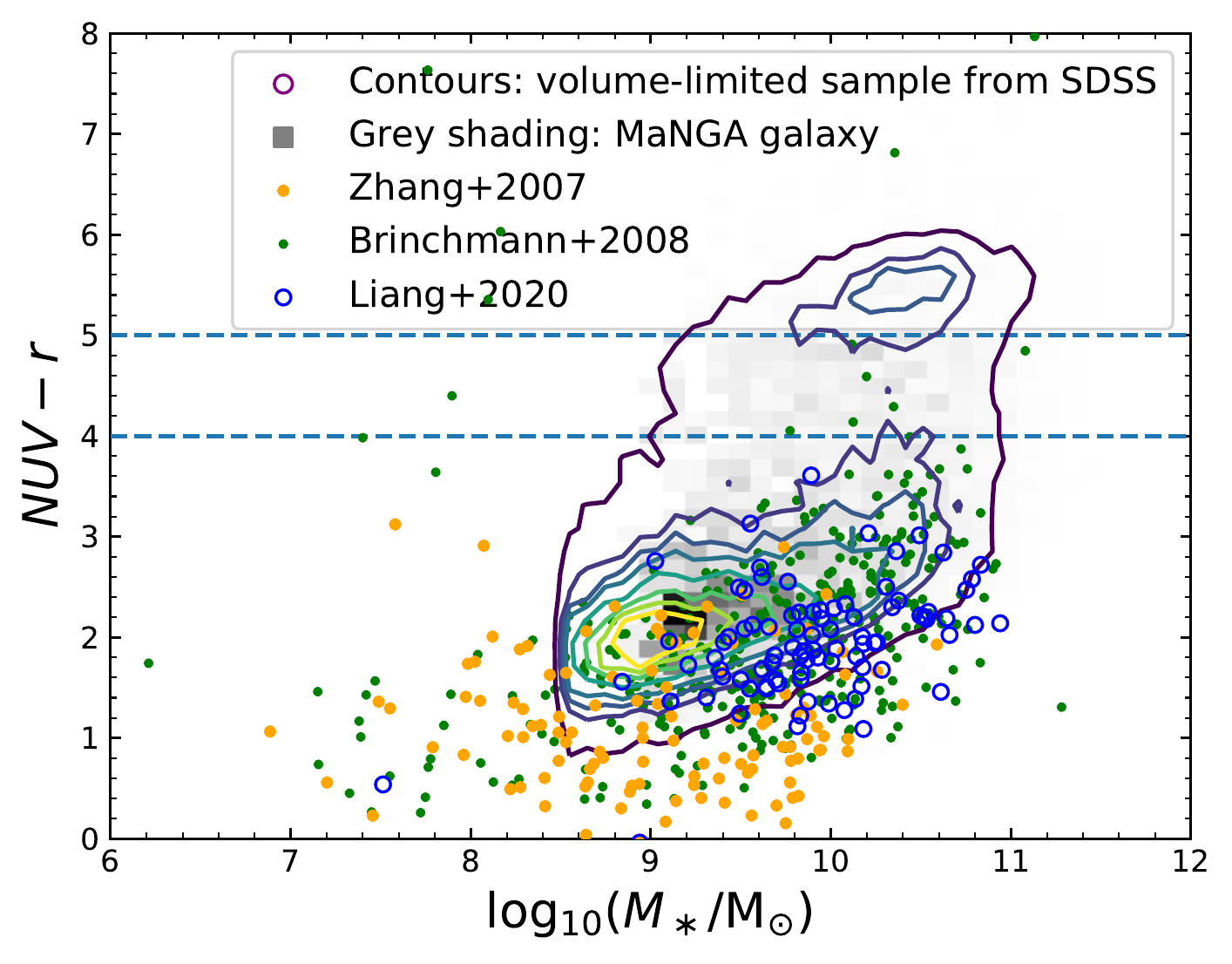}
    \caption{The color-mass diagram of the combined WR sample. We use $NUV-r$ color and stellar mass both based on elliptical Petrosian photometry from NSA catalog. The contours and the gray shading both represent the general population of galaxies. Contours show the number density of galaxy points on this diagram of a volume-limited sample selected from NSA catalog ($z<0.03$, $M_\ast>10^{8.5}\mathrm{M_\odot}$, $h=0.7$). Gray shading is MaNGA galaxies corrected by volume weights provided in \citet{Wake-17} so as for a volume-limited sample. Blue circles are WR galaxies from paper I and orange and green dots are WR galaxies of Z07 and B08 respectively. Cyan lines are empirical boundaries among the blud cloud, the green vallay and the red sequence of galaxies.}
    \label{color}
\end{figure}

\autoref{color} displays the distribution of the WR galaxies 
in our sample on the plane of $NUV-r$ color versus stellar mass 
$M_\ast$. Measurements of $NUV-r$ and $M_\ast$ are taken from 
the {\tt v1\_0\_1} catalog of NASA Sloan Atlas 
\citep[NSA;][]{Blanton-11}, which includes 
SDSS-based measurements of galaxy properties for 
$\sim0.6$ million galaxies at $z<0.15$.
For comparison, the volume-corrected distribution of the 
MPL-7 sample is plotted as the gray background, and a 
volume-limited sample selected from SDSS is shown as the 
contours. The SDSS sample consists of 43,573 galaxies selected 
from the NSA with redshifts $z<0.03$ and stellar masses 
$M_\ast>10^{8.5}M_\odot$. The WR galaxies are plotted 
as colored symbols, with green dots for those from 
B08, orange dots for those from Z07 and blue circles 
for those from Paper I. 

Based on the analyses of completeness of catalogs in Paper I, we now further discuss the physical parameter coverage of the three catalogs in \autoref{color}.
Overall, as expected, the WR galaxies are found exclusively 
in strongly star-forming galaxies with bluest colors at their 
stellar mass. The galaxies from Z07 are systematically bluer 
and less massive than the MaNGA galaxies from Paper I, while 
the B08 catalog covers a wider range in both color and mass 
than the other two catalogs. The differences among the three 
catalogs reflect the selection biases in empirical searching 
for WR galaxies. 
Z07 required significant $EW(\mathrm{H\epsilon})$ emission 
and thus picked the most intense star forming regions. 
Paper I was based on the lower-order Balmer line 
$\mathit{\Sigma}(\mathrm{H\alpha})$ and thus the sample is dominated 
by more general star-forming galaxies, with the rare population 
of the Z07-type galaxies being barely included due to the 
small size of the parent sample. The B08 catalog mainly covers 
a similar area as the MaNGA WR catalog but it also extends to 
the regime of the bluest galaxies, thus it is 
most complete among the catalogs in terms of the global 
property coverage. This may be attributed to two factors: 
the fact that the B08 catalogs did not rely on high-order 
Balmer lines, and the much larger parent sample 
from which the B08 catalog was selected. Although small 
in galaxy number, the MaNGA-based catalog is unique for 
the many off-center WR regions that were missing in the 
SDSS-based catalogs. 

\subsection{Measuring the WR blue bump}
\label{fitting_bump}

For each of the 910 spectra in our catalog, we have performed 
full spectral fitting and measured the flux and equivalent 
width of the WR blue bump based on the starlight-subtracted 
spectrum. Our spectral fitting code was originally developed 
by \citet{Li-05} and further improved in Paper I. In short, 
we construct stellar templates by applying the technique 
of principle component analysis (PCA) to the {\tt MILES}
single stellar population (SSP) models \citep{Vazdekis-10,Vazdekis-15}. 
The first nine eigenspectra produced by the PCA are adopted 
as fitting templates, and are used to fit the SDSS spectrum 
(for the WR galaxies from Z07 or B08) as well as the stacked spectrum of 
the WR region (for the WR regions from MaNGA). During the 
fitting, all significant emission lines as well as the wavelength 
range of the WR blue bump (4600-4750\AA) are masked out. 
Apart from the continuum, we also obtain emission line measurements such as $F_{\mathrm{H\beta}}$ and $EW_{\mathrm{H\beta}}$ in this step.
Details can be found in \citet{Li-05} and Paper I about the
construction of the stellar templates and the scheme of 
masking emission lines. 

For comparison of WR features with model predictions, 
it is important to reliably measure the profile of the WR bump, 
which is a combination of WR star-driven emission lines 
plus nebular emission lines that are irrelevant to the WR feature.
Basically, the blue bump consists of the following WR features: 
\ion{He}{2} $\lambda$4686, 
\ion{N}{5} $\lambda \lambda$ 4605, 4620, \ion{N}{3} 
$\lambda \lambda \lambda$ 4628, 4634, 4640 and 
\ion{C}{3}/\ion{C}{4} $\lambda \lambda$ 4650, 4658. 
The nebular emission lines falling in the WR bump include: 
\ion{He}{2} $\lambda$4686 (narrow line), 
[\ion{Fe}{3}] $\lambda \lambda \lambda$4658, 4665, 4703, 
[\ion{Ar}{4}] $\lambda \lambda$4713, 4740, \ion{He}{1}  
$\lambda$4713, [\ion{Ne}{4}] $\lambda \lambda$4713, 4725. 
Given the current spectral quality, it is not possible to accurately fit so many features independently. 
Based on \citet{Brinchmann-Kunth-Durret-08} and \citet{2014_NGC_3310, 
2016_Califa_WR}, and the occurrence of each 
feature and their relative strength, we have determined the 
following fitting procedure as well as initial guesses and fitting ranges of each parameter.

Starting from the residual spectrum of the full spectrum fitting, 
we first perform a linear fit to the two side windows 
at 4500-4550 {\AA} and 4750-4800 {\AA}, and calculate the root 
mean square ($RMS$) in the two windows. Then, we fit the following five 
features, each with a Gaussian: the \ion{He}{2} bump centered at 4686 {\AA}, 
the C and N bump at around 4645 {\AA}, the \ion{He}{2} 
$\lambda$4686 narrow line,  the [\ion{Fe}{3}] $\lambda$4658 narrow line, 
and the [\ion{Ar}{4}] $\lambda$ 4713 and \ion{He}{1} $\lambda$ 4713 narrow line. 
During the fitting the Gaussian center of the \ion{He}{2} bump and narrow lines are 
fixed at the known wavelength, while the center of the C and N 
bump is allowed to change, but limited to the range of 
$\lambda=4645\pm7.5$ {\AA}. 
Gaussian width $\sigma$ of narrow lines are fixed at the value of 
$\sigma_\mathrm{H\beta}$ while that of broad components are
free parameters within the range $0-22${\AA}.
This upper limit of 22 {\AA} corresponds to the velocity 
dispersion  (1425 km/s) of the WR wind, and this width is almost the largest 
possible value according to literatures on individual WR stars \citep[e.g.][]{2004ApJS..154..651W}.
The linear fit obtained in the first step is used as the baseline 
in this multi-Gaussian fitting and the $RMS$ is used to estimate the 
significance of emission features. 
We keep the bumps or lines if the flux from their Gaussian fit 
has a significance larger than 3.  Next, if the width of the C and N bump 
around 4645 {\AA} reaches the upper limit of 22 {\AA}, 
another Gaussian is added to fit the broad component at 4612 {\AA}. 
Finally, if the peak flux of the residual spectrum (with all the above 
fitted components removed) is larger than 4 times the $RMS$, 
a new Gaussian is added in the fitting, which is either 
a broad component if the peak occurs at $<4650$ {\AA}
or otherwise a narrow comopent. The step is iterated until all fluxes 
in the residual spectrum is lower than 4 times the RMS or at most three additional components are added in this step. 
The fitting procedure uses the python package {\tt PyAstronomy}
\citep{pya}\footnote{\url{https://github.com/sczesla/PyAstronomy}}.

%During the fitting, the Gaussian $\sigma$ for all the narrow lines  
%is tied to that of the H$\beta$ emission line, while $\sigma$ of 
%broad components is set free but limited to the range of 0-22 {\AA}. 

\begin{figure*}[t]
    \centering
    \includegraphics[width=\textwidth]{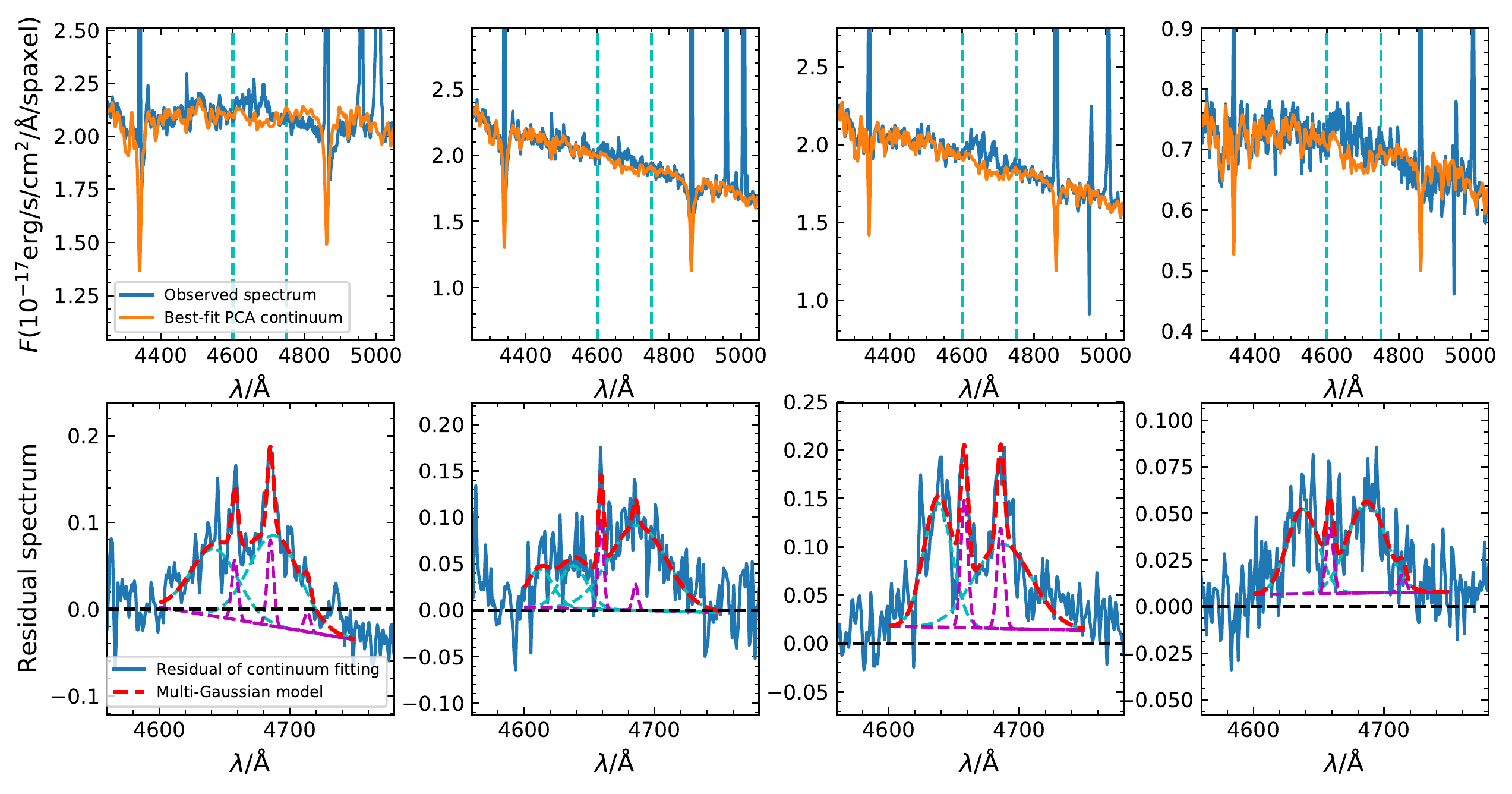}
    \caption{Four examples of the fitting to the WR blue bump. The upper panels show the observed WR spectra (in blue) and their full spectral fitting models (in orange) from PCA algorithm with {\tt MILES} templates. The lower panels show the blue bump feature in the residual and zoom-in the WR window shown with the two vertical dashed lines in the upper panels. The two (or three) cyan lines are the C/N bump(s) (at $\sim 4645 \mathrm{\AA}$) and the \ion{He}{2} bump (at 4686\AA) while the purple lines are narrow nebular emission lines. The red line is the superposition of all components. }
    \label{bump-fitting}
\end{figure*}

\begin{figure*}
    \includegraphics[width=1.0\textwidth]{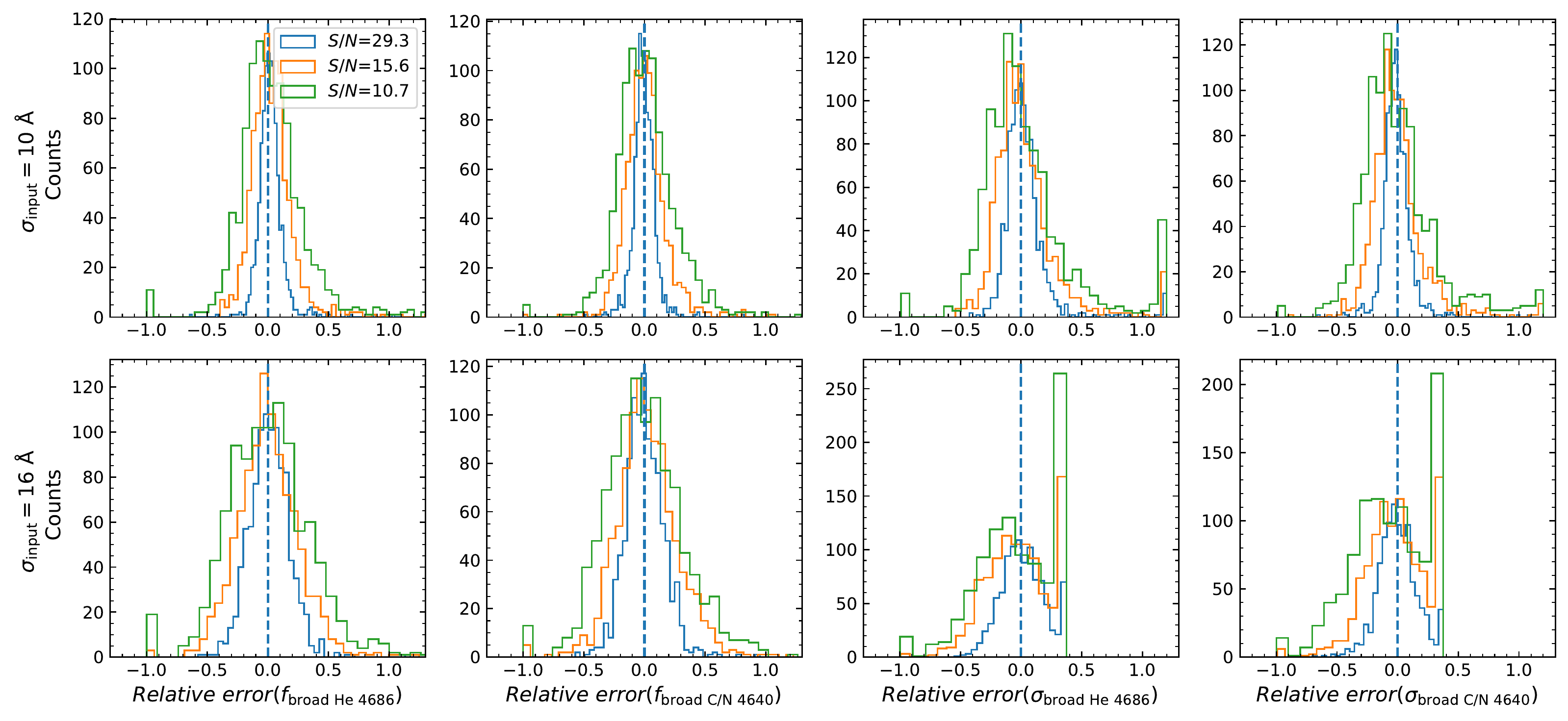}
    \caption{Simulation of the multi-Gaussian fitting to artificial WR blue bumps. We add random noise to artificial WR bumps and calculate relative error between the measured values from fitting and the input values. We perform this procedure for 1000 times for each input combination ($\sigma_{\mathrm{in}}$, broad emission $S/N$) and show the distribution. The upper panels have input Gaussian width of $\sigma = 10${\AA} (for both the C/N and the \ion{He}{2} bumps) while in the lower panels the two bumps each has width of $\sigma = 16${\AA}. Different colors in each panel correspond to different broad emission $S/N$. In four columns, from left to right, we show the distributions for four measured parameters of the two broad emission lines: flux of the \ion{He}{2} bump, flux of the C/N bump, width of the \ion{He}{2} bump and width of the C/N bump.}
    \label{fitting-simulation}
\end{figure*}

\autoref{bump-fitting} shows an example of the Gaussian fits. 
The two (or three) Gaussians plotted in cyan are the C/N bump(s) around 4645 {\AA}
and the \ion{He}{2} bump at 4686 {\AA}, while the magenta 
curves are the narrow nebular emission lines. 
The red curve is the superposition of all the fitted components.

In order to test the reliability of our fitting. We have generated a set of 
mock spectra by artificially creating a WR blue bump with two different 
widths (Gaussian $\sigma$=10 {\AA} and $\sigma$=16 {\AA}) and including three 
different levels of uncorrelated Gaussian noise. Each spectrum includes 
two broad components (the C/N bump at 4645 {\AA} and the \ion{He}{2} 
bump at 4686 {\AA} with equal flux and equal width of either 10 {\AA}  or 16 {\AA} for different mock sets) and three narrow emission lines 
(\ion{He}{2} $\lambda$4686, [\ion{Fe}{3}] $\lambda$4658, 
[\ion{Ar}{4}] $\lambda$ 4713 and \ion{He}{1} $\lambda$ 4713). We fit the 
mock spectra 
in the way described above and repeat the process for 1000 times for each mock bump with newly generated noise. In \autoref{fitting-simulation}, we show 
the distributions of the resulting relative error for the flux and width of the 
two broad components. We can see that, for \ion{He}{2} and C/N bump fluxes, 
the relative difference between the input and output values is centred
at zero, and shows a Gaussian-like distribution with a FWHM that 
ranges from $\sim20\%$ at broad emission $S/N\sim30$ to $\sim 50\%$ at $S/N\sim10$. 
For the width of the two bump components, the overal distribution of 
the relative error is similar to that of the flux, but 
an apparent piling up is seen at the right-hand side, with stronger 
effect at lower $S/N$. This is a sign of the fitting hitting the upper 
boundary of the width before convergence probably due to noise. 
We check the fits for the real spectra and find such case indeed happens 
in some spectra. We visually examined these spectra and 
found they are fitted as well as other spectra. We understand this 
widening in the integrated WR emission as an effect of non-zero velocity 
dispersion of the targeted WR stellar population, for the adopted upper 
limit in fitting is based on individual WR stars. We also note that 
our results to be presented in the next section would not change if 
these cases were excluded from the analysis.

      We should also mention that besides the blue
      bump, the red bump of broad emission lines around 5800 {\AA} is
      also an important feature of extragalactic WR population
      \citep[e.g.][]{Crowther-07} and potentially sensitive to IMF
      variation \citep[e.g.][]{Guseva-Izotov-Thuan-00,
        2002A&A...394..443P}. However, with current sensitivity of our
      data, detectable red bump features are only in a small fraction
      of WR regions. The occurrence and completeness of red bumps in
      our sample are rather small, mentioned in Paper I. Thus, we do
      not measure the red bump or compare it with model
      predictions. It is worth noticing that, however, the red bump
      (if existing) is indeed taken into account in our second
      approach which constrains the IMF models based on full spectral
      fitting (to be elaborated in \autoref{bigs} and
      \autoref{sec:direct}), thus including all features that are
      sensitive to IMF variation. 

\subsection{Starburst99 model}

\citet{1999ApJS..123....3L} publicized the online stellar evolutionary 
model {\tt Starburst99}, based on the \citet{1995ApJS...96....9L} models. 
It incorporates a set of evolutionary model of different stars from 
zero-age main sequence to supernova explosion. It simulates the 
evolution of stellar parameters including mass, luminosity, effective 
temperature, surface chemical composition, etc and assign stellar spectra 
to obtain the synthesized total spectrum at each time slice given different 
physical conditions such as metallicity and the IMF. This model is 
designed for starburst or young stellar population. It is thus among 
the few models that incorporate WR population. Its WR feature is 
essentially the same as described in \citet{Schaerer-Vacca-98}. 

The advantage of {\tt Starburst99} is the very fine time grids and a wide 
range of IMF slopes. WR population is very young and short-lived. 
Its lifetime is typically less than 10 Myr, which is by order of magnitudes 
smaller than normal stellar population. {\tt Starburst99} calculates stellar 
evolution by 0.1 Myr precision and thus provides 90 steps in the range 
of 1-10Myr. It provides complete sets of predicted WR feature 
parameters with IMF slope $\alpha = 1.00,2.35,3.30$ and stellar metallicity 
$Z=0.0004, 0.001, 0.004, 0.008, 0.02 (\mathrm{Z_\odot}), 0.04$.

Besides different physical conditions, {\tt Starburst99} also allows for 
different star formation histories (SFHs). For $\alpha = 2.35$, it provides 
model predictions with five different SFH \textemdash\ instantaneous 
burst and constant star formation of 1, 2, 10, 100 Myr.

\subsection{BPASS model}

In addition to {\tt Starburst99}, we also use the latest stellar evolution 
model {\tt BPASS v2.2.1} \citep{2018MNRAS.479...75S, 2017PASA...34...58E}, 
which provides models for both singular stellar populations and binary 
stellar populations. This allows us to examine the effect of binary populations
on WR features as well as the resulting constraints on the IMF slope. 
The effect could be significant. For instance, when binary evolution is 
considered, the lifetime of WR population can be extended from 5 Myr to 10 
Myr \citep{2009MNRAS.400.1019E}. {\tt BPASS} adopts atmosphere models
from Potsdam PoWR group \citep{2003A&A...410..993H, 2015A&A...577A..13S} 
and a distribution of binary parameters including binary fraction
presented in the Table 13 of \citet[][]{2017ApJS..230...15M}, which were based on 
local stellar populations and lack information for sub-solar metallicity 
populations as a caveat.

{\tt BPASS} has finer metallicity grids than {\tt Starburst99} but a narrower 
IMF slope range and coarser time grids. It provides complete sets of 
single stellar population spectra with 13 different metallicities from 
$Z=1.0\times 10^{-5}$ to $Z=0.040$, and three IMF slopes 
$\alpha = 2.00,2.35,2.70$. it has two sets of models with an upper 
mass limit of $M\mathrm{_{upper}} =100\mathrm{M_\odot}$ and $300\mathrm{M_\odot}$. 
The time resolution is 0.1 dex and therefore has 
only 10 grids in 1-10 Myr. Unlike {\tt Starburst99}, {\tt BPASS} only provides 
instantaneous bursty models.

{\tt BPASS} simple stellar population (SSP) models involve WR stellar 
emission lines when WR stars present. In order to compare the strength 
of the WR bump between model and data, we have measured the equivalent 
width of the blue WR bump of the SSP's, adopting a local linear baseline. 
The WR bump exists only in very young populations so the stellar absorption 
lines are not abundant and the stellar continuum is rather flat (except 
for the oldest snapshot at $\log_{10}(Age)=7.5$ of $Z=0.02$ or higher, 
to be mentioned later). In the next subsection (\autoref{bigs}), we will also use these SSP models 
as templates to directly fit the spectra of WR galaxies/regions in our sample. This allows direct modeling of WR spectra with SSP's. 

\subsection{Bayesian inference of WR spectra}
\label{bigs}

In \autoref{fitting_bump} we have performed full spectral fitting 
to each of the 910 WR spectra in our sample, in order to separate 
emission lines (particularly the WR blue bump) from the stellar 
spectrum so that the WR bump can be reliably measured. Now 
we perform full spectral fitting to the WR spectra again, but using 
{\tt BIGS} (Bayesian Inference for Galaxy Spectra), which is a 
Python spectral fitting code developed by \citet{2019MNRAS.485.5256Z}. 
{\tt BIGS} adopts the approach of Bayesian inference and Markov chain 
Monte Carlo algorithm to model the composite stellar populations 
(CSP) present in the observed WR spectra. The code incorporates 
different stellar population models and all IMF's discussed in previous subsections. By comparing the Bayesian evidence for different models,
one can make model selection reliably and efficiently. In \autoref{sec:direct}
we will take advantage of this virtue to compare {\tt BPASS} 
models of different IMF's. 

The working process of {\tt BIGS} can be summarized as follows. 
First, a stellar IMF and a specific star formation hisotry (SFH) 
are assumed. The SFH model is then combined with 
a set of SSP templates of {\tt BPASS} and a simple screen 
dust model \citep{Charlot2000} to generate composite model spectra, 
which are then convolved with the stellar velocity dispersion derived 
above (\autoref{fitting_bump}) to account for kinematic and instrumental 
broadening effects. Next, the model spectra are compared with each of 
the observed spectra, and accordingly the likelihood for a given set 
of model parameters is calculated. The {\tt MULTINEST} samplper 
\citep{Feroz2009,Feroz2013} is utilized to sample the posterior distributions 
of the model parameters and derive the Bayesian evidence.  
The {\tt BIGS} code has been successfully appiled to MaNGA data
to constrain the IMF slope for ellitpical galaxies \citep{2019MNRAS.485.5256Z}, 
as well as SFHs of both low-mass galaxies \citep{Zhou2020a} and 
massive red spiral galaxies \citep{Zhou2020b}. 

We have fitted each of the WR spectra using {\tt BIGS}. Specifically, 
the WR population is modeled by a recent nearly-instantaneous burst 
with lookback time 0-10 Myr , later referred to as the ``bursty stellar population". This component is superposed to an 
underlying old stellar population modelled by the commonly-adopted 
exponential $\tau$-decaying model.  Free parameters in fitting are dust attenuation quantified 
by color excess $E(\mathrm{B-V})$, burst-to-total 
mass ratio $f$, stellar metallicity of the bursty stellar population $Z_\mathrm{b}$, standard parameters of the $\tau$ model (i.e. stellar metallicity $Z_\tau$, lifetime $\tau$, and starting time $t_0$).
The two metallicities $Z_\mathrm{b}$ and $Z_\tau$ are fitted independently in the fitting. 
We take the best-fit metallicity of the bursty stellar population as the metallicity of 
the WR population (for simplicity, we define $Z\equiv Z_\mathrm{b}$ hereafter), and use it 
for all following figures and discussion. We note that we have dropped 
a few spectra with unreasonably low metallicity in the tail of the 
derived $Z$ distribution 
($\log_{10}(Z/\mathrm{Z_\odot})<-1.5,\ \mathrm{Z_\odot} = 0.02$). 
During the fitting we have masked out emission lines in a iterative 
manner \citep[see][for details]{Li-05}, but the WR bump is kept and 
fitted. 

%\autoref{bigs_example} shows the spectral fit obtained with {\tt BIGS} for 
%one of the WR spectra as an example. It is encourating to see that the
%WR features, both the broad component and the narrow lines, are well 
%reproduced in the best-fit spectrum. We have carefully masked out 
%pixels with problematic data as indicated by the grey vertical bands 
%in the bottom panel. 

\begin{figure}
    \includegraphics[width=\columnwidth]{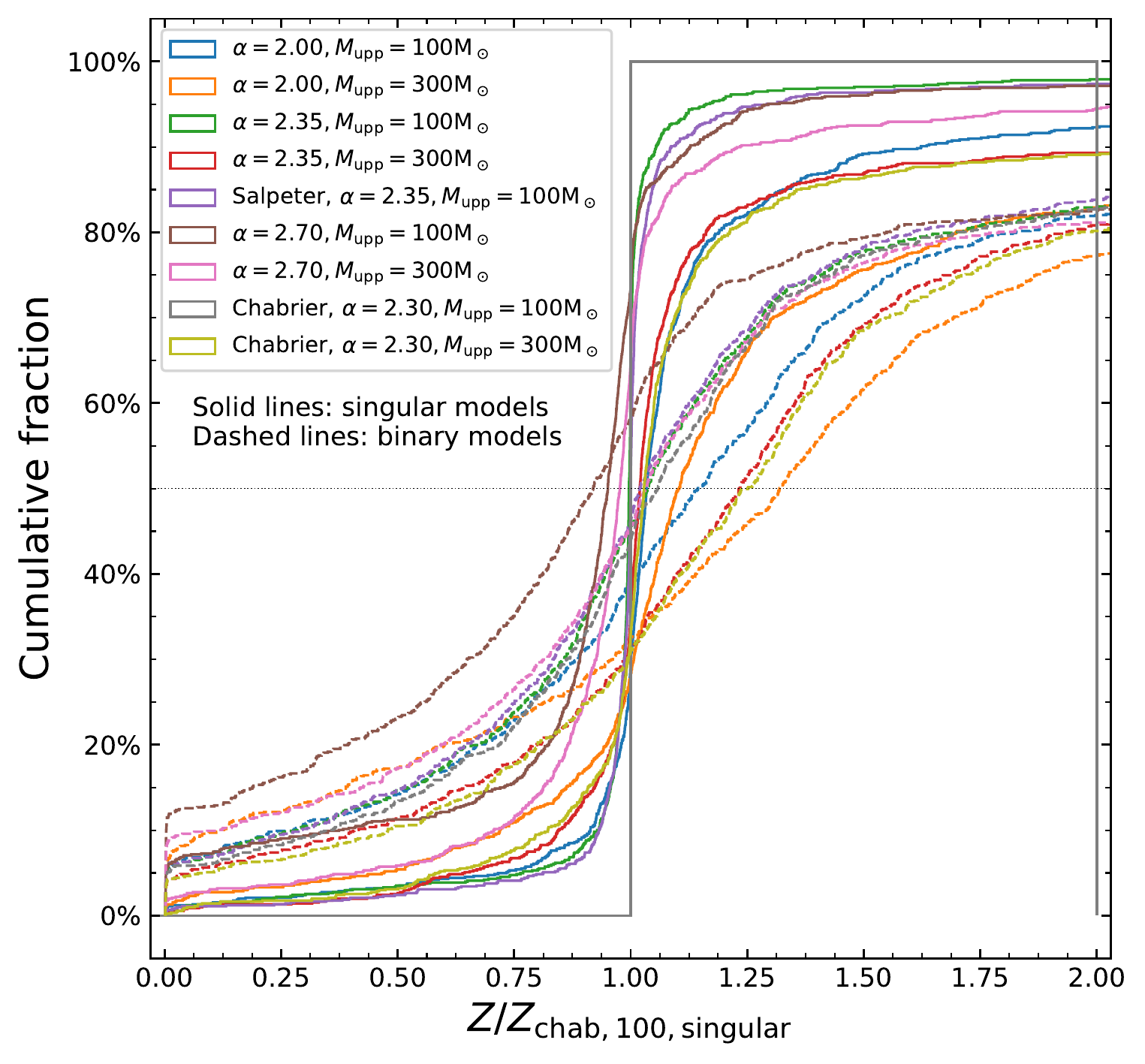}
    \caption{The comparison of derived bursty metallicity among different IMF assumptions. Each WR spectrum has a set of derived $Z$'s from fitting with different IMF assumptions. We arbitrarily set the singular model of Chabrier IMF with 100 M$_\odot$ upper limit to be the standard and divide all other metallicity values of each spectrum by its $Z_{\mbox{chab,100,singular}}$. We then plot the cumulative distribution of the ratios. The gray line of Chabrier IMF with 100 M$_\odot$ upper limit is, by construction, entirely 1.00. The horizontal dotted line crosses each cumulative distribution at its median value.}
    \label{bigs_z_all_lin}
\end{figure}

For each spectrum, we have done multiple fittings using all the available
SSP templates from {\tt BPASS}. These include two sets with different types of evolution (i.e. singular and binary), and each
set consists of nine SSP models with different IMF assumptions (corresponding to four different
high-mass slopes paired with two different upper % ($\alpha=2.00$, $2.30$, $2.35$,$2.70$)
mass limits, plus an additional Salpeter IMF; see \autoref{tab:model} for details).  % ($M_{\mbox{upper}}=100$, $300\mathrm{M_\odot}$) 
As a result,  for each WR spectrum, we have 18 fittings and correspondingly 18 derivations for each aforementioned physical quantity such as  $E(\mathrm{B-V})$, $Z$, etc. 
\autoref{bigs_z_all_lin} shows the cumulative distributions
of the measurements of the stellar metallicity (the bursty stellar population), with solid (dotted)
lines for the singular (binnary) population models and different
colors for the  different IMF's, as indicated. We have scaled the $Z$
measurements of each WR spectrum by the measurement of an
arbitrarily chosen model,  which is the singular model with a Chabrier
IMF and an upper mass limit  of 100 M$_\odot$. Overall, as one can
see, different models give rise to similar distributions of $Z$,
as indicated by the sharp rise at around
$Z/Z_{\mbox{chab,100,singular}} = 1$ and the relatively narrow range  of
the median $Z$ (indicated by the horizental dotted line) spanned  by
the models.  The largest difference between binary models and this particular singular model, which is $\substack{+0.31\ \mathrm{dex}  \\ -0.06\ \mathrm{ dex}}$, occurs
with the binary model of $\alpha=2.70$ and
$M_{\mbox{upper}}=100\ \mathrm{M_\odot}$  (the dotted brown line) and the binary
model of $\alpha=2.00$ and  $M_{\mbox{upper}}=300\ \mathrm{M_\odot}$ (the dotted
orange line). The differences  are smaller if the comparison is
limitted to the singular population models only, with the larggest
difference ($\lesssim20\%$) occuring between  the same pair of
models. Apart from looking at the median values, the smaller differences between the singular  models can also
be seen from the more sharply rising of the distributions.  These
results indicate that the $Z$ measurements are weakly dependent  on
the adopted IMF's, but there are systematic differences betweeen  the
singular and binary models for a given IMF. In what follows, we will
need to divide our WR galaxies/regions into subsets according to
stellar  metallicity. For simplicity, we use the median measurement of
$Z$ for each WR galaxy/region, but separately for the case of singular
and binary population models. 

\begin{figure*}
    \includegraphics[width=\textwidth]{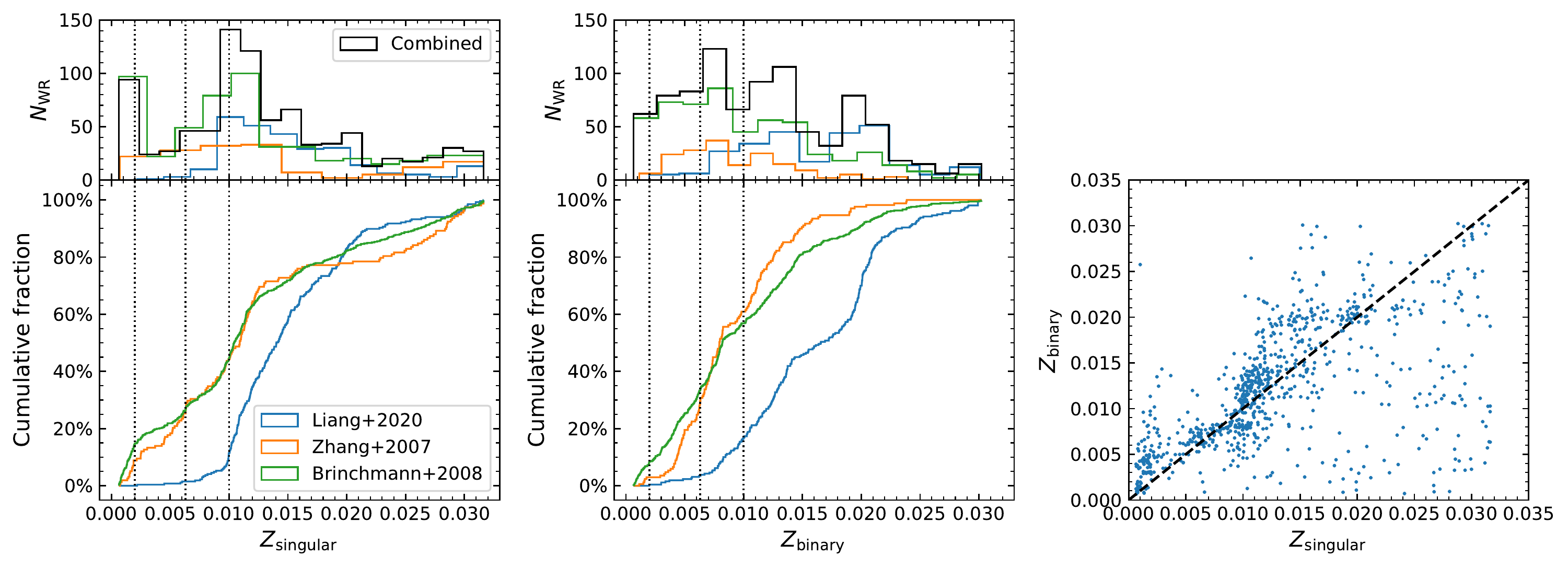}
    \caption{The distribution of the combined WR sample in $Z$. The histogram (number of WR spectra or regions $N_{\mathrm{WR}}$) and cumulative distribution of $Z$ derived from sigular (binary) model fitting is shown in the left (middle) panel. Different colors indicate different WR catalogs. See \autoref{bigs} for details in derivation of $Z$. 
    % Note that for each WR spectrum and assuming singular (binary) stellar evolution, {\tt BPASS} provides 9 sets of model SSPs, with 9 corresponding different IMF's; we 
    % We use all {\tt BPASS} templates with 18 different IMF's in BIGS fitting. 
{Vertical dotted lines show the boundaries of our four $Z$ bins used in \autoref{imf_zhang} and \autoref{ew_age}. We set the bin boundaries according to the availability of $Z$ grids in {\tt Starburst99} models and {\tt BPASS} models.}
     In the right panel, we show the scatter plot of $Z_{\mbox{singular}}$ and $Z_{\mbox{binary}}$. When Z07 and B08 are plotted individually, they both have their original full catalogs while in the combined distribution (black histograms) and the scatter plot (right panel), their common 101 galaxies only appear once.}
    \label{sample_z_distribution}
\end{figure*}

\autoref{sample_z_distribution} displays the  distribution of the median
stellar metallicity mesurements for the singular population models
(left panel) and the binary population models (middle panel).  The
main panels show the cumulative distribution functions, while the upper panels
show the corresponding histograms. Results for the three WR
catalogs are shown separately. The overal distribution  of all the WR
galaxies/regions in our sample is plotted as the black histogram in
the upper panels. 
When Z07 and B08 are plotted individually, they both have their original full catalogs while in the combined distribution (black histograms) and the scatter plot (right panel), their common 101 galaxies only occur once.
Our sample spans a wide range of metallicity (from Z$\sim$0.001 up to Z$\sim$0.03),  as
one can see from the figure. The Z07 and B08 catalogs present similar
distributions in $Z$, while the WR catalog from MaNGA is relatively
metal-rich.  This might be implying that the WR regions at galactic
centers as selected by  Z07 and B08 are more metal-poor than those in
the disk which dominate the  WR catalog selected from MaNGA. We will
come back to this point in a later paper of this series (Liang et al. in
prep.). The vertical dotted lines in the left and the middle panels indicate the
boundaries used to select subsamples in \autoref{sec:parameter} for \autoref{imf_zhang} and \autoref{ew_age}. {We set the bin boundaries according to the availability of $Z$ grids in {\tt Starburst99} models and {\tt BPASS} models.}
 The right panel compares the median measurements for the singular and
binary population models. Most data points distribute along or closely
to the identity line, while a considerable fraction of the data points
deviate from the line. This means the choice of using singular evolution model or binary evolution model can cause a significant difference in the derived metallicity to a small fraction of observed spectra. This further suggests it is necessary to
consider the effect of binary populations when modelling the
underlying  stellar populations of observed spectra. 

{ In \autoref{tab:model}, we summarize key information of our modeling described in this section.
}

\begin{table*}[]
%\color{blue}
\caption{A table summarizing all models.}
\label{tab:model}
\hspace{-30pt}
\begin{tabular}{llccc}
\hline
\hline
\multicolumn{1}{c}{}                                                                                     & \multicolumn{1}{c|}{}                                                                                          & \multicolumn{2}{c}{Singular modeling}                                                                                                               & Binary modeling                                                       \\
\multicolumn{1}{c}{}                                                                                     & \multicolumn{1}{c|}{}                                                                                          & {\tt Starburst99}                                                                  & {\tt BPASS} singular                                                       & {\tt BPASS} binary                                                          \\ \hline
\multicolumn{1}{l|}{\multirow{4}{*}{\hspace{-35pt}  \begin{tabular}[c]{@{}l@{}}Model\\ feature\end{tabular}}}            & \multicolumn{1}{l|}{IMF Slope $\alpha$}                                                                                & 1.00, 2.35,  3.30                                                             & 2.00, 2.30, 2.35, 2.70                                               & 2.00, 2.30, 2.35, 2.70                                                \\
\multicolumn{1}{l|}{}                                                                                    & \multicolumn{1}{l|}{IMF upper cutoff $M_{\text{upper}}$ in M$_\odot$}                                                                              & 120                                                                          & 100, 300                                                             & 100, 300                                                              \\
\multicolumn{1}{l|}{}                                                                                    & \multicolumn{1}{l|}{Metallicity $Z$ grids}                                                                        & \hspace{-30pt}\begin{tabular}[c]{@{}c@{}}4E-4, 1E-3, 4E-3, \\ 8E-3, 0.02 0.04\end{tabular} & \hspace{-30pt} \begin{tabular}[c]{@{}c@{}}13 grids from\\ 1E-5 to 0.04\end{tabular} &\hspace{-30pt} \begin{tabular}[c]{@{}c@{}}13 grids from \\ 1E-5 to 0.04\end{tabular} \\
\multicolumn{1}{l|}{}                                                                                    & \multicolumn{1}{l|}{Number of age grids in 0-10 Myr}                                                                    & 90                                                                           & 10                                                                   & 10                                                                    \\ \hline
\multicolumn{1}{l|}{\multirow{4}{*}{\hspace{-35pt} \begin{tabular}[c]{@{}l@{}}Employment\\ in this paper\end{tabular}}} & \multicolumn{1}{l|}{Use as SSP templates in BIGS fitting}          & No                                                                           & Yes                                                                  & Yes                                                                   \\
\multicolumn{1}{l|}{}                                                                                    & \multicolumn{1}{l|}{Use in $Z_{\text{WR}}$ derivation}                                                                  & No                                                                           & Yes, $Z_{\text{singular}}$                                                          & Yes, $Z_{\text{binary}}$                                                             \\
\multicolumn{1}{l|}{}                                                                                    & \multicolumn{1}{l|}{Use in WR property comparison in §3.1}         & Yes                                                                          & Yes                                                                  & Yes                                                                   \\
\multicolumn{1}{l|}{}                                                                                    & \multicolumn{1}{l|}{Use in Bayesian inference of IMF slope in §3.2} & No                                                                           & Yes                                                                  & Yes                                                                   \\ \hline
\end{tabular}
\end{table*}

\section{Constraining the high-mass slope of the IMF}
%\subsection{Compare data and model prediction}
\label{sec:results}

\subsection{Constraining the IMF slope with SSP's}
\label{sec:parameter}

\begin{figure*}
    \includegraphics[width=\textwidth]{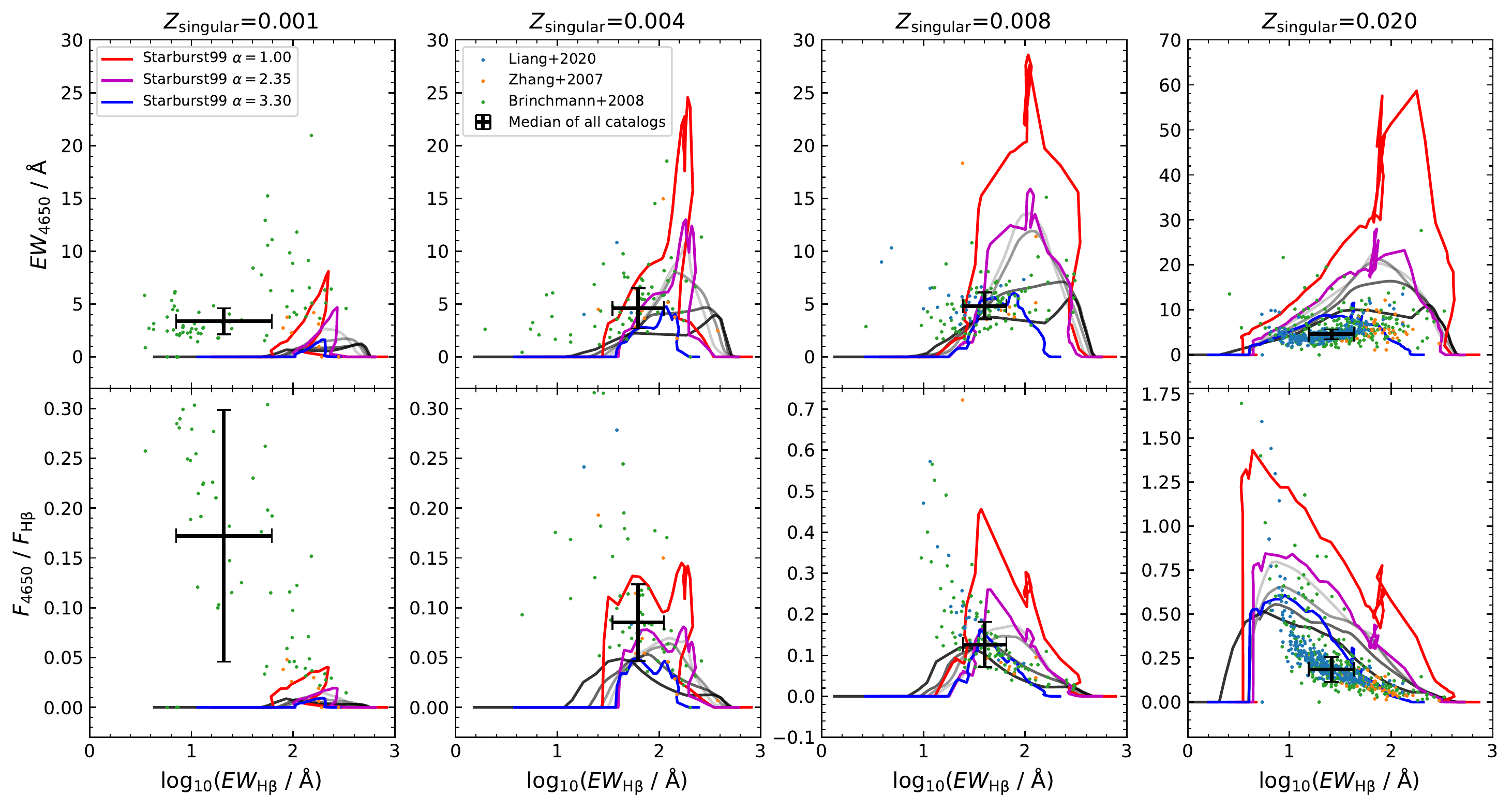}
    \caption{We compare WR parameters of the observation and the singular 
    evolutionary model (namely {\tt Starburst99}) in this figure. The upper four 
    panels show comparison in the plane of $EW_{4650}$ vs log($EW_{\mathrm{H\beta}}$) 
    while the lower four panels show  $F_{4650} / F_{\mathrm{H\beta}}$ vs 
    log($EW_{\mathrm{H\beta}}$).
    WR regions (shown as dots) are divided into four $Z$ bins (four different columns) and compared with 
    corresponding model predictions (shown as evolutionary tracking lines). 
    The large black cross is the median value of data points 
    in each panel with errorbars showing median absolute deviation. Different 
    colors of lines are different IMF assumptions with the red line being 
    the steepest IMF slope ($\alpha=3.30$) and the blue line being the 
    flattest slope ($\alpha=1.00$). The gray lines are assuming the same 
    IMF slope as the megenta line ($\alpha=2.35$) but assuming more extended SFH than colored lines of instantaneous starburst. 
    The darker the gray, the more prolonged the SFH.}
    \label{imf_zhang}
\end{figure*}

First, following Z07, we examine the correlation of IMF slope with
stellar metallicity by comparing the observed WR features in our
sample  with that of the SSP's from {\tt Starburst99}, using WR
galaxies/regions  in different stellar metallicity intervals. The
results are shown in  \autoref{imf_zhang} where we plot the
distribution of $EW_{4650}$ (upper panels) and $F_{4650} /
F_{\mathrm{H\beta}}$ (lower panels) versus $EW_{\mathrm{H\beta}}$. The
measurements of $EW_{4650}$ and  $F_{\mathrm{H\beta}}$ are obtained in
\autoref{fitting_bump}, while  $F_{\mathrm{H\beta}}$ and
$EW_{\mathrm{H\beta}}$ are described in  \autoref{sec:catalog}. As
pointed out in Z07, the use of $EW_{4650}$ and  ratio to $\mathrm{H\beta}$
can alleviate the uncertainties from dust extinction.  In the figure,
panels from left to right correspond to four metallicity  ranges: $Z <
0.002$, $0.002 < Z < 0.0063$, $0.0063 < Z < 0.01$, and $Z > 0.01$.
The median metallicity of the observed galaxies/regions in each bin is $Z=0.0013$, 0.0045,
0.0089, 0.0150 in case of singular models, and $Z=0.0012$, 0.0043,
0.0079, 0.0152 in case of binary models. In each panel the small
points  represent WR galaxies or regions from our combined sample, and the
big cross and the error bar indicate the median of the data points
and the median of the absolute deviation from the data's median.
We use the median of data points and the median absolute deviation,
instead of the mean average and the standard deviation as often
adopted, in order to minimize the effect of outlier data points. 

For comparison, predictions of evolutionary tracks of instantaneous
starbursts from {\tt Starburst99} with metallicity of $Z=0.001$,
0.004, 0.008, 0.020 are plotted as colored lines in the four panels.
In each panel, lines of different colors represent the three different
IMF slopes: $\alpha=1.00$, 2.35, 3.30.  As one goes from low to high
metallicities, the location of the data points  move significantly
with respect to the model lines. As a result, models  of different IMF
slopes are needed to interpret the data at different metallicities.
In the lowest metallicity bin ($Z=0.001$), all the data points are far
beyond the enclosed parameter range of all the models.  At $Z \sim 0.004$,
quite a fraction of the data points as well as the median
value fall in between the red ($\alpha=1.00$) and magenta
($\alpha=2.35$) lines.  At $Z\sim0.008$, more data points fall below
the magenta line, and the median  of the data points fall slightly
below the blue line  ($\alpha=3.30$). Finally, in the highest
metallicity bin where $Z=0.02$,  the majority of data points and their
median value are well enclosed by the model with  the steepest IMF
(blue line, $\alpha=3.30$). These clearly suggest  a
metallicity-depdent IMF slope in the sense that the IMF steepens with
increasing metallicity. From the figure, as one can see, this trend is
attributed to the significant change of the model tracks  with
metallicity in contrast to the weak metallicity-dependence of the
distribution  of the data points in the two parameter diagrams. In all
panels,  the median of $EW_{4650}$ remains $\sim$5{\AA}, while the
median of  $F_{4650} / F_{\mathrm{H\beta}}$ spands a limited range of
$0.12-0.18$.  In the models, however, for a fixed IMF slope both
$EW_{4650}$  and $F_{4650} / F_{\mathrm{H\beta}}$ change significantly
with   metallicity. Taking the megenta line ($\alpha = 2.35$) for
example,  the peak value of $EW_{4650}$ increases from $\sim$5{\AA} at
$Z=0.001$  to $\sim$28{\AA} at $Z=0.02$, and the peak value of
$F_{4650} / F_{\mathrm{H\beta}}$  increases from $\sim0.02$ to
$\sim0.85$. Combining the  observational (lack of) trend and the
theoretical trend, one can clearly  conclude that any single IMF
cannot explain the data at all metallicities, and that a
``bottom-heavy'' (``top-heavy'') IMF is needed to interpret  the data
at high (low) metallicites. This result is well consistent with  Z07
and recent studies which use different catalogs/data and methods
\citep[e.g.][]{2012Natur.484..485C, 2017ApJ...838...77L,
  2018MNRAS.477.3954P, Zhou2020a}.

\begin{figure*}
    \includegraphics[width=\textwidth]{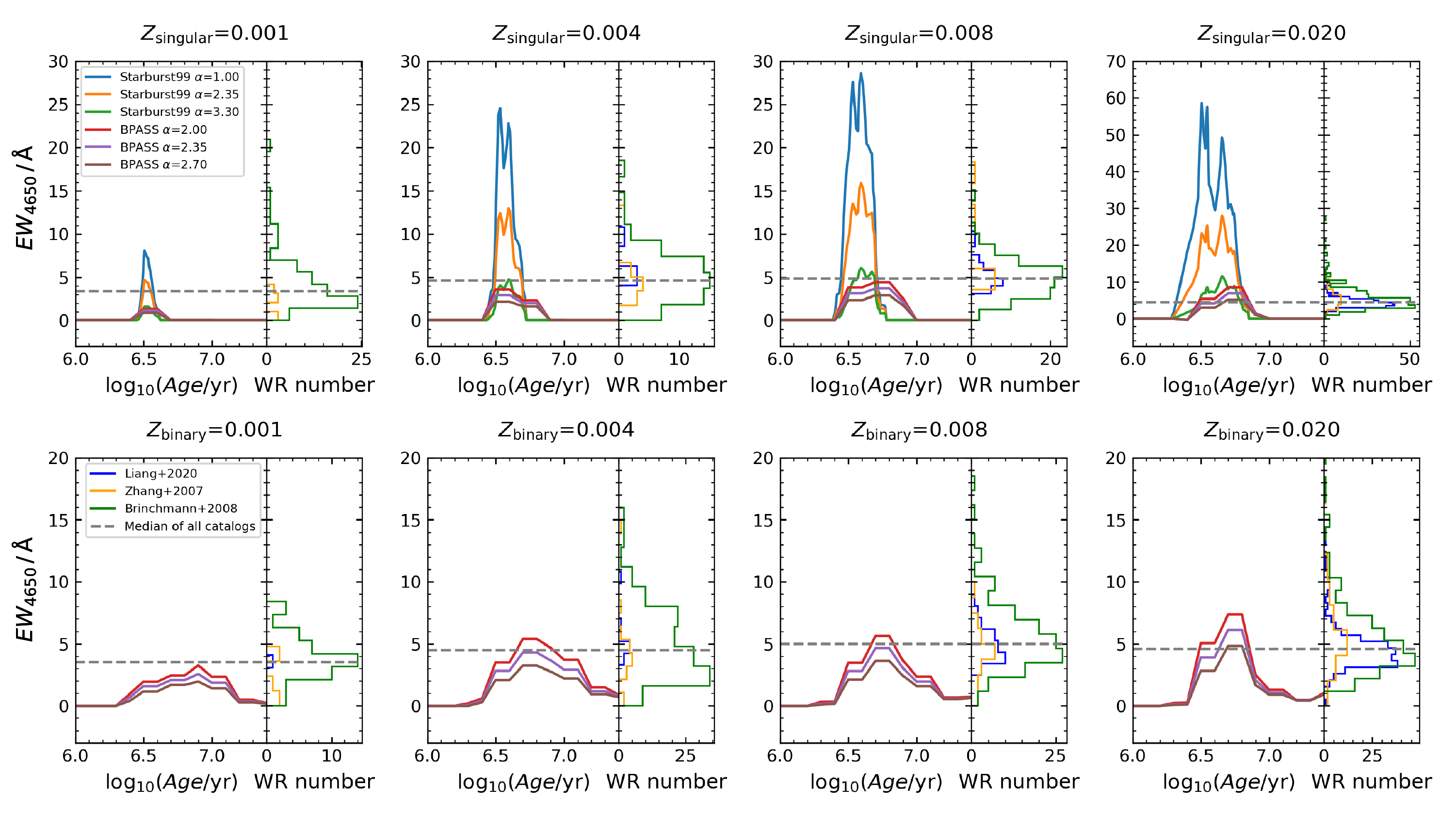}
    \caption{The evolution of WR parameters in singular evolution models
    (upper panels) and binary evoluiton models (lower panels). Different colunms 
    are for different $Z$ bins. In each panel, the main sub-panels on the left 
    show model predictions from {\tt Starburst99} and {\tt BPASS} with different 
    colors indicating different IMF's, and the side panels on the right show 
    the histograms of observed $EW_{4650}$ of WR spectra in respective $Z$ bins. The 
    horizontal dashed lines is the median value of the combined WR sample in each 
    $Z$ bin. Note that the slight rise of model lines towards $\log_{10}(Age) = 7.5$ is an artifact due to linear fitting on {\tt BPASS} SSP's where absorption lines prevail.}
    \label{ew_age}
\end{figure*}
%  where the linear baseline fitting to BPASS SSP models is driven down by an increasing influence of metal absorption lines at this highest $Z$ bin. In other words, the increase does not come from WR feature

In addition to the instantaneous burst models, we also consider
model  predictions from {\tt Starburst99} for star formation with
longer durations.  We show the results in \autoref{imf_zhang} for the
Salpeter IMF ($\alpha=2.35$) as different shades of gray lines in each
panel. The four lines  from light gray to dark gray  correspond to
SFHs with duration $t=1,2,10,100$ Myr. As the burst duration
increases, the predicted WR bump becomes weaker when compared to
$\mathrm{H\beta}$ emission. This can be explained by the different
lifetimes between  the WR population (which produce the WR feature at
around 4650\AA) and other OB stars (which produce the
$\mathrm{H\beta}$ emission).  For an instant starburst of given IMF
and metallicity, WR stars and OB  stars are formed at the same time
and mixed instantaneously, and the WR  feature is relatively strongest
in this case. For an extended burst, since  WR stars have shorter
lifetimes than other OB stars, at any given time  the existing WR
population is mixed with both contemporary OB stars and  those fromed
at earlier times. This dilutes the feature of WR stars against the
continuum (the denominator of $EW_{4650}$) and the $\mathrm{H\beta}$.  In
\autoref{imf_zhang}, we can see gray lines basically span the
parameter  space between the megenta lines and the blue lines. Thus,
variation  in SFH alone cannot fully interpret the data when the IMF
slope  is fixed at $\alpha=2.35$.

\begin{figure*}
    \includegraphics[width=\textwidth]{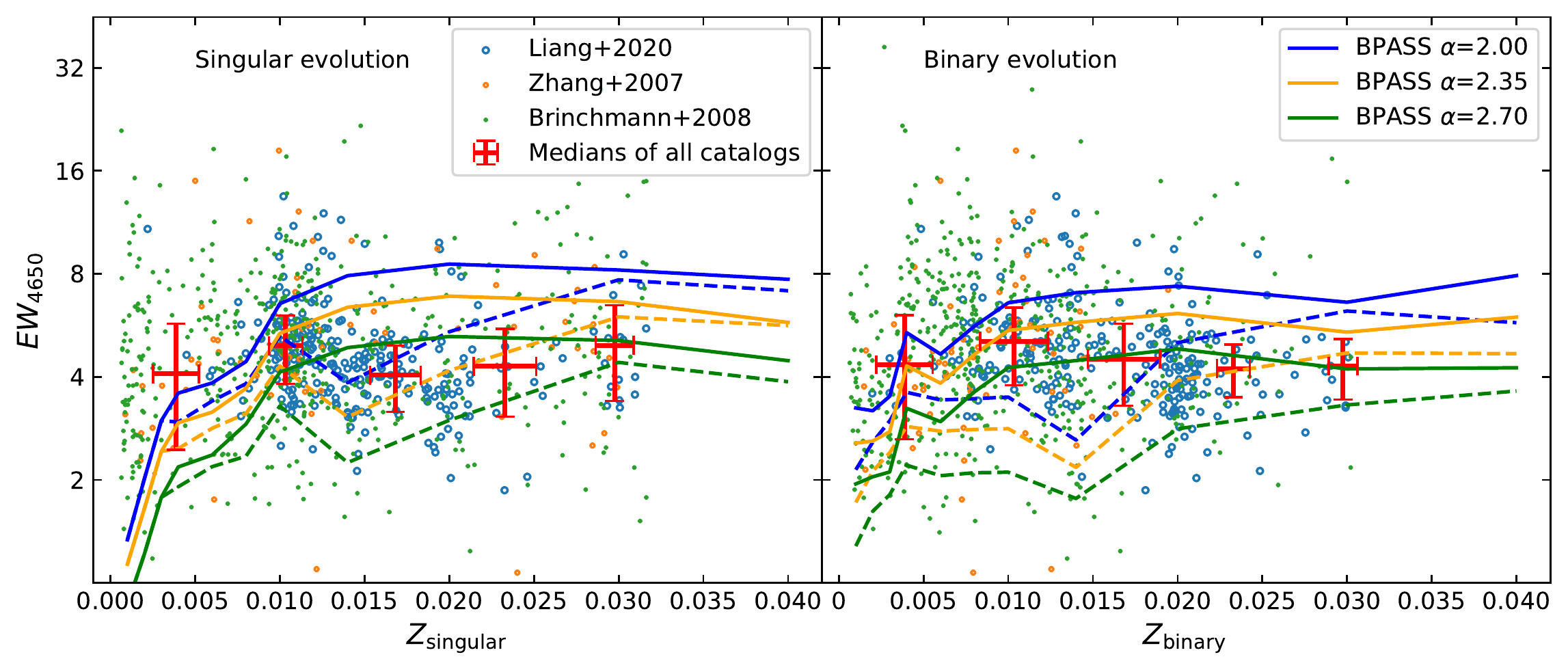}
    \caption{Comparison of $EW_{4650}$ and $Z$ for both singular evolution (left) and binary evolution (right). Different colored lines are model predictions with different IMF slopes. { Since it is impossible on this 2D figure to show another dimension of time evolution of $EW_{4650}$, we instead show the maximum values during the WR phase with solid lines and the median values of $EW_{4650}$ (among snapshots when significant WR feature exists) with dashed lines.} Red crosses are median value of the data points at different $Z$ bins. Errorbars are median absolute deviation.}
    % We use the maximum values of $EW_{4650}$ from all snapshots of model predictions with a given IMF, for WR population spend their majority of time at around peak value $EW_{4650}$, as seen in \autoref{ew_age}.
    \label{ew_z}
\end{figure*}

Next, in \autoref{ew_age}, we show the evolution of the WR parameter
$EW_{4650}$ with time as predicted by {\tt BPASS} for both singular
populations  (upper panels) and binary populations (lower panels), and
for different  IMF slopes (different lines) and metallicities
(different columns). For the singular populations, the predictions  of
{\tt Starburst99} are also plotted for comparison. From the
observational side, we have obtained measurements of $EW_{4650}$ for
all the WR  galaxies and regions. However, it is hard to reliably estimate
ages for such young stellar populations, considering the age-metallicity degeneracy and uncertainties in the age derived from stellar population synthesis. Also, the star formation history we use in \autoref{bigs} does not set the age of the starburst as a free parameter. Therefore, we plot the
histograms of $EW_{4650}$ of the WR galaxies/regions in side panels,
with  the median value indicated by the horizontal dashed lines.  As
one can see from the upper panels, although the stellar ages  from the
two models cover similar ranges, the WR features from {\tt
  Starburst99}  are much stronger than that from {\tt BPASS} for given
IMF and metallicity. 
%Compared to {\tt Starburst99}, the {\tt BPASS} predictions  appear to better agree with our data. 
Despite the discrepancy in the absolute value of $EW_{4650}$, the two
models  predict the same trend that, at fixed IMF slope, $EW_{4650}$
increases  with increasing metallocity. Since the median value of
$EW_{4650}$  in the sample depends very weakly with metallicity, both
models  requires an increase in IMF slope as one goes from low to high
metallicity in order to bring the mdoels better match the data.  For
instance, the model of {\tt Starburst99} with $\alpha=2.35$  matches
the data in the lowest metallicity bin, while the model  with
$\alpha=3.30$ is most compatible to the data in the highest
metallicity bin. In case of {\tt BPASS}, all the models are much
lower than the data in the lowest metallicity bin, which may be
attributed to the limited range of IMF slope in the model
($\alpha=$2.00-2.70). Models of $\alpha=2.00$ get closer to the data
in the two intermediate metallicity bins, while the model of
$\alpha=2.70$ best matches the data in the highest metallicity bin.

In the lower panels, the need of steeper IMF's at  higher metallicities
can also be seen for the binary population models.  In the lowest
metallicity bin, the model with the flatest IMF  ($\alpha=2.00$)
matches the median of the sample, while in the  two intermediate
metallicity bins the data falls in between  the models with
$\alpha=2.00$ and $\alpha=2.35$. In the highest  metallicity bin, it
is the model with the steepest IMF ($\alpha=2.70$) that best matches
the data.  This result suggests that the  metallicity-dependent
variation of the IMF slope holds in  both singular population models
and binary population models. 

% added
The difference caused by varying between $M\mathrm{_{upper}}=100\ \mathrm{M_\odot}$ and $M\mathrm{_{upper}}=300\ \mathrm{M_\odot}$ (with other parameters fixed) is tiny. Thus, for simplicity we do not differentiate this upper mass limit, where relevant.

In \autoref{ew_z} we plot $EW_{4650}$ as a function of stellar
metallicity, for both the data and model predictions. We consider {\tt
  BPASS}  models of both singular (left panel) and binary (right
panel) populations.  
%For given IMF and metallicity, we use the peak
%value of $EW_{4650}$  in the model, since in \autoref{ew_age} we have
%seen that the  rise and fall of predicted WR feature is very rapid
%with most time  staying around the peak value. 
Galaxies/regions from
our sample are plotted in small dots, while the red crosses  and
erorbars represent the median and the median absolute deviation in  a
given metallocity bin. 
{  For given IMF and metallicity, we show both the peak value of $EW_{4650}$ (with solid lines) as well as the median value (with dashed lines) among snapshots when significant WR feature exists (more specifically, the median value among SSP's with at least 10\% of the peak value, which is a reasonable threshold considering the observed $EW_{4650}$ range). }
% By this means, we qualitatively address the uncertainty of the ages of our WR population.
Overall, as already seen from previous figures,
the median observed $EW_{4650}$ shows weak dependence on metallicity in the
data.  In the models, with fixed IMF slope the predicted $EW_{4650}$ increases %{peak } 
with increasing metallicity, and the effect is most remarkable  at
lowest metallicities ($Z\lesssim 0.01$). At higher metallicities, the
predicted  $EW_{4650}$  flattens out.  %{  peak }
Combining the weak varation  (or no variation) of the observed $EW_{4650}$ at all $Z$'s and the significant increase of
the model-predicted $EW_{4650}$ as a function of $Z$ assuming a fixed IMF, one can expect  that we need different IMF assumptions at different $Z$'s in order to reproduce observed values at all $Z$'s.  More specifically, we need to assume  
a flatter IMF (i.e. ``top-heavy'') at lower metallicities and a steeper IMF
(i.e. ``bottom-heavy'') at higher metallicities.A more detailed comparison between data and model is following, which leads to this IMF variation in a more quantitative way.

When comparing observed data points with models in \autoref{ew_z}, one needs to compare the data points with the intervals spanned between peak values and median values at any given IMF.
%{ For the predicted {\it median} $EW_{4650}$, the general trend of rising with $Z$ for any given IMF is the same as the predicted {\it peak} values, but more fluctuation is seen in intermediate $Z$ bins. 
Since the ages of observed WR regions are unknown, different WR regions should be at randomly different phases of WR lifetime. Thus, observed WR sample should have a distribution (rather than an exact value) of $EW_{4650}$ at any given $Z$ bin. Besides, the detection completeness of WR regions is higher for regions with higher $EW_{4650}$, since these bumps are less likely to be destroyed by observational noise in the spectra. Combining the above analysis, we expect 
the detected WR $EW_{4650}$ should mostly fall in this interval (between medians and peaks of model prediction).
In the upper panel, the median observed $EW_{4650}$ shown with red crosses fall in (or are closer to) the blue interval alone ($\alpha=2.00$) at low $Z$ while they fall in the green interval alone ($\alpha=2.70$) at high $Z$. In other words, it is clear a universal IMF does not exist to encompass WR observations at all metallicity. In the right panel of binary modeling, this trend also exists but is less clear. The red cross at the lowest $Z$ falls in the blue interval ($\alpha=2.00$) but is also at the boundary (peak) value of the orange interval ($\alpha=2.35$), which means the flattest IMF slope is slightly preferred in this metallicity. The orange interval also encompasses the red crosses in the three intermediate $Z$ bins. In the highest $Z$ bin, the red cross falls below the orange interval and is at the boundary (peak) of the green interval ($\alpha=2.70$). Thus, the steepest IMF is slightly preferred at the highest $Z$. 
It is interesting that
the same trend holds in both singular and binary population models,
although the metallicity dependence of the WR featuer in the binary
models is slightly weaker than that in the singular models.

% added
It has been seen that for a given IMF slope, models predict stronger WR emission (higher $EW_{4650}$) at higher metallicity. This is due to the positive correlation between metallicity and the mass loss rate (and therefore stellar wind emission) of WR stars \citep{Crowther-07}. The absence of such trend in observed WR population means at lower metallicity, more WR stars are needed to compensate for their relatively weaker stellar winds than at high metallicity. This immediately implies a ``top-heavy'' IMF at low metallicity.

We should stress that all these analysis should be considered in a statistical manner. In other words, we always compare the average of data points with models. If we examine individual data points, we can see many of them locate well apart from any model prediction, especially at the low metallicity end. The discrepancy at low metallicity is very large for singular evolution case, e.g. in the bottom-left panel of \autoref{imf_zhang}, the upper-left panel of \autoref{ew_age}, and the leftmost part of the left panel of \autoref{ew_z}. The introduction of the binary evolution models alleviate this discrepancy, which should be a demonstration of the importance of binary evolution. Nonetheless, even when we look at the binary case, some individual points still deviate from all model predictions. This can be attributed to uncertainties in measuring WR bumps, the relatively narrow range of available IMFs in BPASS models, lack of knowledge in high-mass stellar evolution at low metallicity, etc. We leave deeper investigation of the discrepancy between models and individual WR regions to future studies.

\subsection{Bayesion inference of the IMF slope from full spectral fitting}
\label{sec:direct}

\begin{figure}
    \includegraphics[width=\columnwidth]{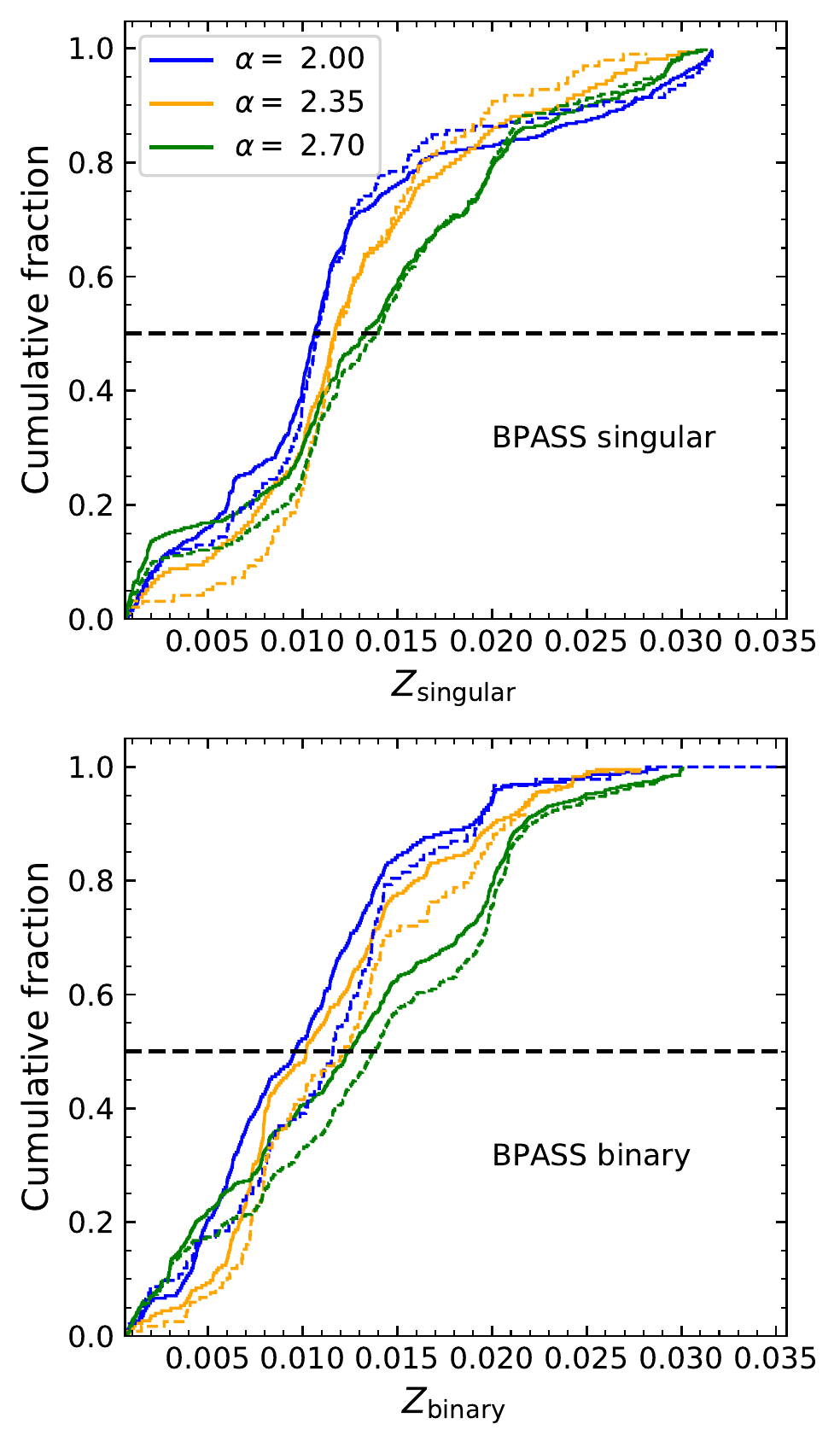}
    \caption{Cumulative distribution of $Z$ for WR sample grouped by their optimal IMF slope determined from Bayesian evidence of fitting. The upper panel is the singular fitting result and the lower panel is the binary result. Colored dashed lines are high spectral $S/N$ (at $\sim 4675\,  \mathrm{\AA}$) subsamples of corresponding solid lines. Horizontal black dashed lines cross the colored lines at their median values.}
    \label{bigs_slope}
\end{figure}

\begin{figure}
    \includegraphics[width=\columnwidth]{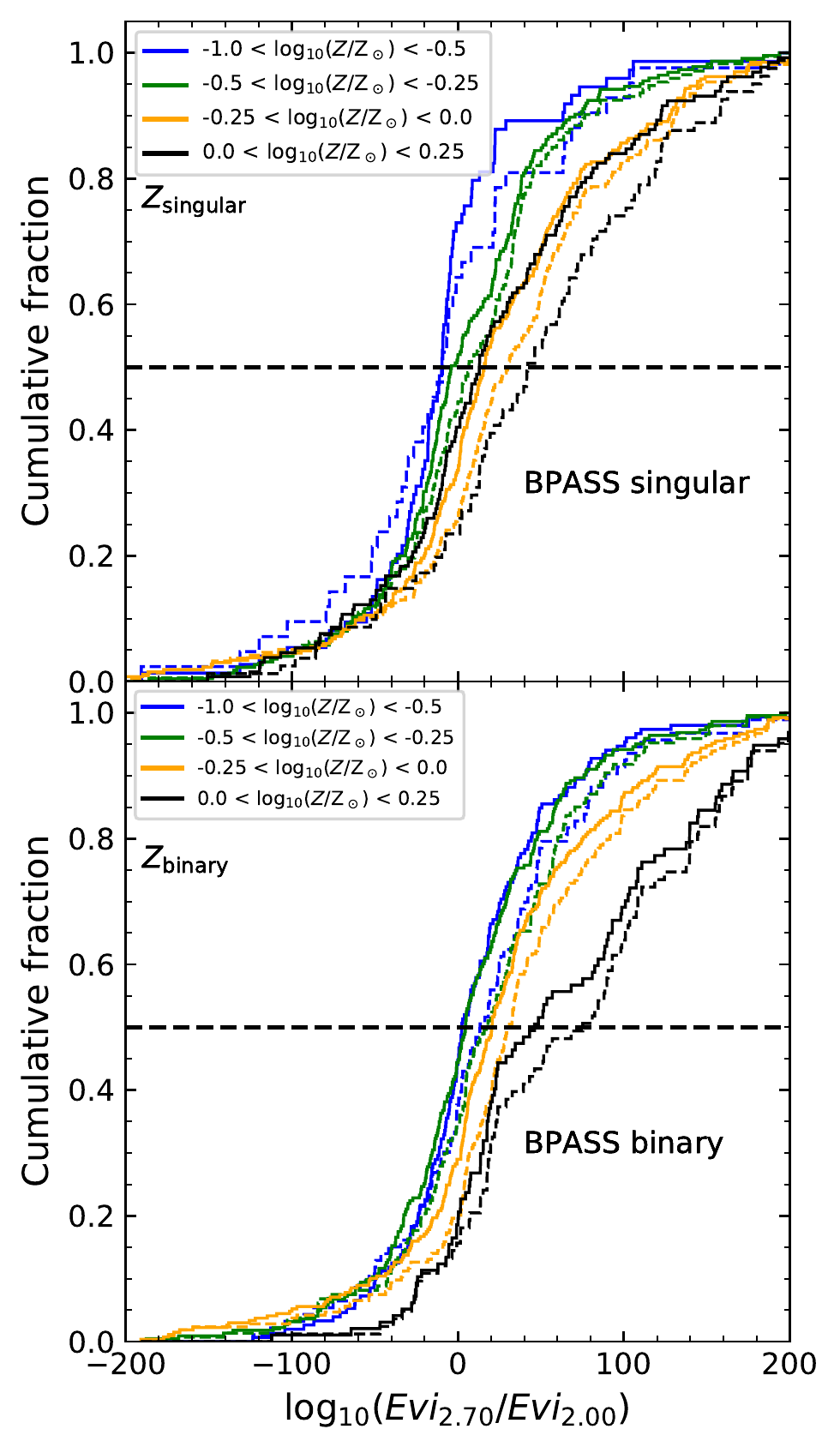}
    \caption{Cumulative distribution of Bayesian evidence ratio
$\mathrm{log}_{10}(Evi_{2.70} / Evi_{2.00})$ for sub-samples grouped by $Z$. For each WR spectrum, since we have fitted it multiple times with different IMF assumptions, we now extract the Bayesian evidence of fitting with $\alpha=2.70$ IMF (i.e. $Evi_{2.70}$) and with $\alpha=2.00$ IMF (i.e. $Evi_{2.00}$) and take the ratio to indicate the preference between the two IMF slopes.
The upper panel is from fitting with singular templates while the lower panel is from binary templates. Colored dashed lines are subsamples of corresponding solid lines with high $S/N$ (at $\sim 4675\,  \mathrm{\AA}$). Horizontal black dash lines cross the colored distribution lines at their median values.}
    \label{bigs_ratio}
\end{figure}

As described in \autoref{bigs}, we have performed full spectral
fitting to each of the WR spectra in our sample using  {\tt BIGS},
which provides Bayesian evidence for us to select  the best model
among many others. We consider the {\tt BPASS} models of both singular
and binary populations. For each spectrum,  we fit it with {\tt BIGS},
using models of the three different IMF slopes: $\alpha=2.00$, 2.35
and 2.70. We then select one of the three  IMF slopes which leads to
the highest Bayesian evidence as the optimal IMF slope for the
spectrum. In this way we divide the 910 WR  galaxies/regions into
three groups according to the optimal IMF slope.  \autoref{bigs_slope}
displays the cummulative distribution of $Z$ for the
three groups. The upper and lower panels show results (solid lines)
for the singular and binary models separately. 
In both panels, the metallicities of the group with the steepest IMF ($\alpha=2.70$,
green lines) are higher than those of the other two groups, as can be
seen from the separation between the distributions as well as the median metallicities indicated by the horizontal
line. The separation between the distributions is apparent 
at $0.01<Z<0.02$. 
%The two groups with $\alpha=2.35$ and $\alpha=2.00$ show little differences, however. 
We note the curves converge at $Z<0.01$ and $Z>0.02$. We suspect this is possibly due to the narrow range of available $\alpha$ in {\tt{BPASS}}. This narrow range of variation makes model less distinguishable at either very low ($Z<0.01$) or very high ($Z>0.02$) metallicity. This convergence may also be attibuted to lack of binary templates, since we can see in the lower panel the separation among curves are generally larger at the high-$Z$ end. 
We also select a
subset of spectra with relatively high spectral signal-to-noise at $\sim 4675\,  \mathrm{\AA}$ ($S/N>20$) and
plotted the results as dashed lines in the same figure. Results of the high-S/N spectra are similar to those of the full sample with separation being slightly clearer in the binary modeling (lower panel). 

\begin{figure}
\includegraphics[width=\columnwidth]{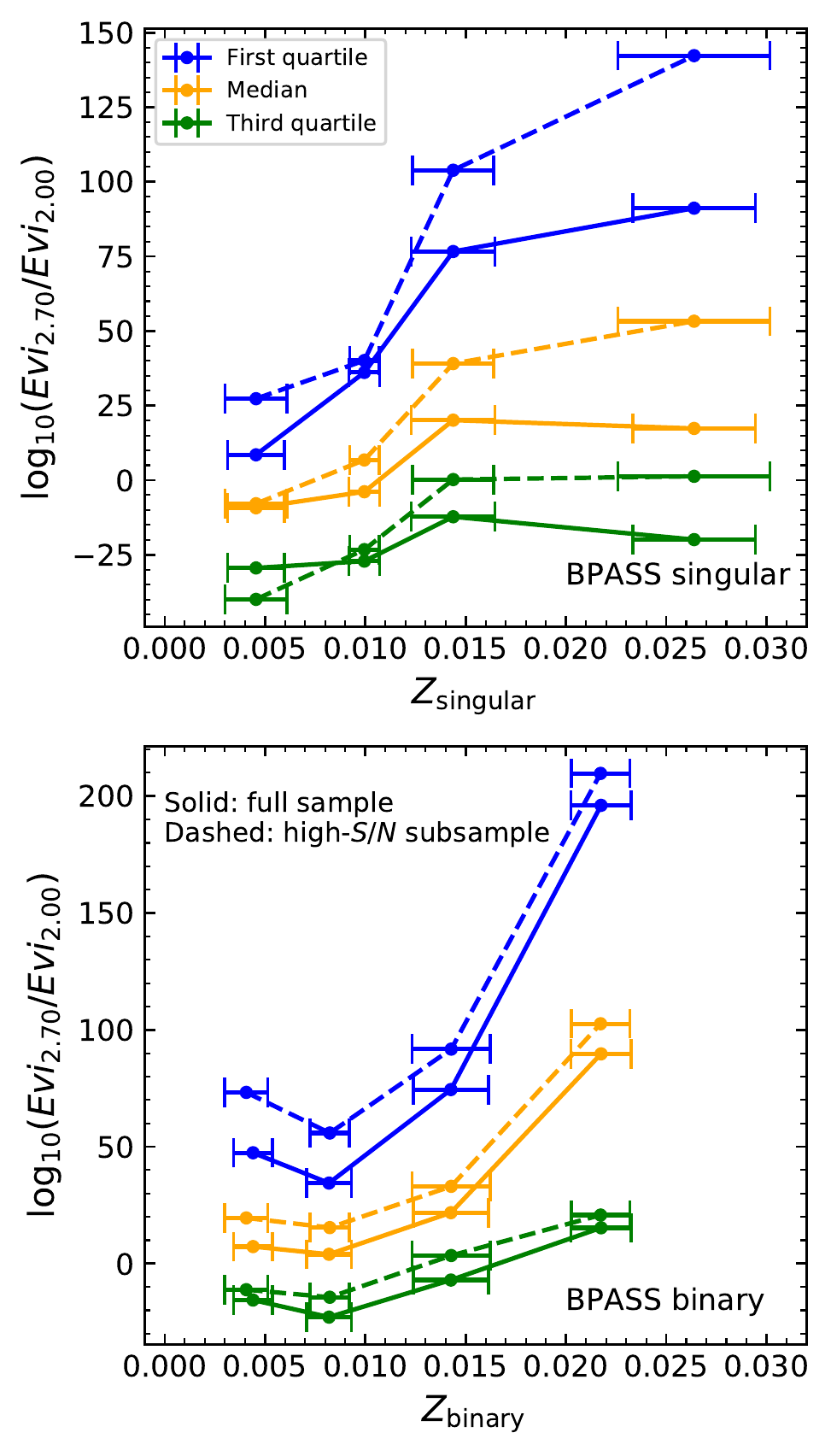}
%\color{blue}
\caption{Evidence ratio $\mathrm{log}_{10}(Evi_{2.70} / Evi_{2.00})$ vs metallicity $Z$. We divide our sample into the same four $Z$ bins as used in \autoref{bigs_ratio}. At each $Z$, we show the three quartiles of evidence ratios in different colors. The solid lines are quartiles of the full WR sample while the dashed lines show those of a high-spectral-$S/N$ subsample. The horizontal error bars are the median absolute deviation of $Z$ in each bins.}
\label{fig:quartile}
\end{figure}

In \autoref{bigs_ratio}, we divide all the WR galaxies/regions into
four bins according to $Z$ specified in the legends. For each spectrum, since we have performed multiple fittings with different IMF assumptions, we now extract the Bayesian evidence from fittings with $\alpha=2.70$ and with $\alpha=2.00$. We calculate the ratio of them to indicate the preference between the two IMF's. For each WR spectrum, if the ratio is larger (or smaller) than one, $\alpha=2.70$ (or $\alpha=2.00$) is favoured. We then
plot the cummulative distribution of the Bayesian evidence ratio $\mathrm{log}_{10}(Evi_{2.70} / Evi_{2.00})$.  Again, we show
results for both the full sample (solid lines) and the subset with
high spectral $S/N$ at $\sim 4675\,  \mathrm{\AA}$ (dashed lines). 
{ Note that the four $Z$ bins in this figure as well as in the following figure are different from those in the previous subsection. In the previous subsection, the bins are set by the availability of $Z$ grids of model SSP's, especially limited by the scarce grids of {\tt Starburst99} model. In this subsection, since we use Bayesian evidence to infer the IMF slope instead of comparing to SSP's, we can better bin our sample by $Z$ into subsamples with roughly equal sizes. Also, we eliminate spectra with $Z<0.002$, for we have shown in the previous subsection that no model matches data well in this lowest $Z$ bin.}
Both the overall distribution 
and the median value of the evidence ratios favor larger
values with increasing metallicity, especially for the subsample of high spectral $S/N$.
This indicates that a steeper ($\alpha=2.70$) IMF is preferred
at higher $Z$. This result holds for both singular and binary models.

{
In \autoref{fig:quartile}, we further present the evidence ratios in a slightly different way from \autoref{bigs_ratio}. We divide our sample into the same $Z$ bins (with dots located at median $Z$ of each bin and median absolute deviation as error bars) and show quartiles of evidence ratios in each bin. 
In general, we can see a positive trend between evidence ratio quartiles (all of the three quartiles) with $Z$ in both panels (both singular case and binary case). In other words, at higher $Z$, the evidence ratios are higher no matter which quartile of the distribution we look at. This again indicates the steeper IMF $\alpha=2.70$ is preferred at higher $Z$, consistent with previous findings.
The high spectral $S/N$ subsample results (shown with dashed lines) present a clearer trend in the singular modeling (upper panel) than the full sample (shown with solid lines) while in the binary modeling (lower panel) they are roughly the same as the full sample. Besides, the trend of singular (binary) modeling exists in lower (higher) $Z$ bins, i.e. $Z\lesssim 0.015$ ($Z\gtrsim 0.007$) and flattens out elsewhere. This means the transition between $\alpha=2.00$ and $\alpha=2.70$ happens at slightly different $Z$ intervals with singular and binary modeling.
%though we note the fluctuation due to different binning can be large (not shown on the plot). 
Besides, when we inspect the values of the evidence ratios instead of the trend, we find the singular modeling experiences a transition from negative median $\mathrm{log}_{10}(Evi_{2.70} / Evi_{2.00})$ to positive median values while the median values in binary modeling are positive at all $Z$. 
% This presumably indicates a degeneracy between binary modeling and IMF variation.
Although this phenomenon does not conflict with the variation of IMF given the clear trend of evidence ratio we have discussed, further studies are needed to better investigate the implied degeneracy between binary modeling and IMF variation.  In addition, evidence ratios at the highest $Z$ bin have a larger variation reflected by the larger separation between the first and the third quartiles at this $Z$.   The origin of this larger scatter is still unknown.
% may be due to the limited IMF range {\color{blue} \textbf{  available in {\tt BPASS} }} and a lack of perfect IMF at this $Z$. 
We also notice the significant difference between median $Z_{\mathrm{singular}}$ and $Z_{\mathrm{binary}}$ in the highest $Z$ bin despite the same binning boundaries. This is a direct result of the systematically different distributions of $Z_{\mathrm{singular}}$ and $Z_{\mathrm{binary}}$.
}

To sum up, we have echoed our findings in \autoref{sec:parameter} by means of Bayesian inference of WR spectra in this subsection. A ``top-heavy'' IMF is preferred in low $Z$ environments. This holds for both singular modeling and binary modeling. A high spectral $S/N$ (at $\sim 4675\,  \mathrm{\AA}$) subsample sometimes shows this trend more clearly than the full sample,
namely in the lower panel of \autoref{bigs_slope}, in \autoref{bigs_ratio} and in the upper panel of \autoref{fig:quartile} .
Nonetheless, we can see limitations of the narrow IMF slope range of {\tt BPASS} models. Models covering wider ranges in $\alpha$ are needed in order to better constrain the relationship between $Z$ and $\alpha$, at both high and low metallicities. 
% We are unable to include flatter IMF's in this analysis due to the limited IMF slope range of the {\tt BPASS} models.  

\section{Discussion}
\label{sec:dis}

\subsection{WR spectra as a unqiue probe of the IMF slope}

WR spectra provide a unique probe to constrain the massive end
of the IMF { with the youngest stellar population ($<10\ \mathrm{Myr}$)} in galaxies. This approach relies on the fact that 
the WR feature in integrated spectra of galaxies or star-forming 
regoins is produced by living massive stars during their WR phase, 
thus sensitive to the IMF slope at the high-mass end. 
This approach is complementary to other probes that have 
mostly focused on old stellar populations in early-type 
galaxies or bulges in late-type galaxies 
\citep[e.g.][]{Cappellari-12,2017ApJ...838...77L,2018MNRAS.477.3954P,2019MNRAS.485.5256Z}.

In an early work, \citet{Guseva-Izotov-Thuan-00} analyzed long-slit
spectra of 39 WR galaxies covering a wide range of gas-phase
metallicity. It was found that, at the lowest metallicites of their
sample, the distribution of galaxies on the diagram of WR bump
equivalent widths versus $EW_{\mathrm{H\beta}}$ can be better (though not fully)
matched by theoretical predictions of the models of
\citet{Schaerer-Vacca-98} if a very shallow IMF with $\alpha=1.00$ is
adopted instead of the Salpether IMF with $\alpha=2.35$.  
Though the original authors did not attribute this phenomenon to IMF variation.
Based on
the SDSS/DR3, \citet{Zhang-07} identified a large sample of 174 WR
spectra and examined both $EW_{4650}$ and $F_{4650}/F_{\mathrm{H\beta}}$ as a
function of $\log_{10}(EW_{\mathrm{H\beta}})$. The authors compared the WR galaxy
distribution on these diagrams, for four different metallicity ranges
separately, with predictions of the single population models of
\citet{Schaerer-Vacca-98} with different IMF slopes (the same models
as used in {\tt Starburst99}). This analysis led to the conclusion
that steeper IMF slopes are needed at higher metallicities in order
for the model to better match the data, in good agreement with the
early hint from \citet{Guseva-Izotov-Thuan-00}. The same trend is also seen
in our work, as shown in \autoref{imf_zhang}, but our work has well
extended the work of \citet{Zhang-07} in the following aspects. 

% {\color{blue} In addition, the approach of modelling WR population allows for determination of metallicity as the more intrinsic driving factor of the IMF variation than other properties such as velocity dispersion of stellar population. } % No, cannot say so. Velocity dispersion is also an intrincis factor related to GMC turbulence.

First, we have significantly enlarged the WR sample and improved the
coverage of galaxy properties for the sample, by combining  the
SDSS-based WR catalogs constructed in previous studies
\citep{Zhang-07, Brinchmann-Kunth-Durret-08} with the MaNGA-based
catalog constructed in our previous work \citep{liang_2020}. As shown
in \autoref{color}, different WR catalogs cover basic galaxy
parameters  in quite different ways, due to different selection
methods.  In addition, the SDSS spectra are limited to central regions
of galaxies, and the MaNGA datacubes include outer regions of galaxies
in our analysis. The combined sample thus includes  a more complete
set of WR regions from different locations  and in differen types of
galaxies. 

Second, for consistency, we have used our own pipeline to re-analyze  all
spectra in the combined WR sample, instead of taking values from
literature. We have carefully decomposed and measured the emission
lines in the WR wavelength window. We have also applied our spectral
fitting code {\tt BIGS} which allows statistical comparison and
selection of different models in a Bayesian inference framework.
These careful treatments on data analysis have provided a solid  basis
to support our conclusions. 

Thirdly, on the theoretical side, we have considered not only the singular evolution model as used in \citet{Zhang-07}, but also the
recently-developed model {\tt BPASS}. The
% removed " It is encouraging that", trying to be more objective
metallicity-dependent IMF slope variation is also supported  by this
new model. In addition, the binary population models in  {\tt BPASS}
have allowed us to examine the effect of binary evolution. At
intermediate-to-high metallicities, we find similar results when
considering binary population models. At the lowest metallicities, models with binary populations
predict larger equivalent widths of the blue WR bump, thus more
closely matching the data than the singular population models. As can
be seen in \autoref{ew_z}, the binary model with the shallowest IMF
well matches the median $EW_{4650}$ of the data at the lowest
metallicities, while all the sigular models significantly underpredict
the same quantity. This is the first time that binary evolution is
explicitly considered in the constraint of the IMF.

\subsection{Systematics and caveats in interpreting the results}

Several sources of uncertainties and random error could have come into our analysis. First, we rely on stellar continuum subtraction in measuring WR features. This can cause some uncertainties in our measured $F_{4650}$ and $EW_{4650}$, due to residual continuum, in addition to the random error due to spectral noise. Second, in the derivation of metallicity of our spectra, we assume a two-component star formation history when carrying out stellar population synthesis. This is another souce of model-dependency. Again, we stress that given these uncertainties, we do not analyse any single WR spectra. All our results are based on a large sample, which should be more robust against uncertainties.

We mainly test two potential systematics. The first is contamination of WR stellar population by other populations, such as old underlying population and diffuse ionzed gas (DIG) surrounding WR pupulations. In principle, our second approach discussed in \autoref{sec:direct} is not affected by this contamination, for we have already included and then subtracted the contribution of old populations to the spectra. However, this can affect our measured $EW_{4650}$ since the old population can contribute considerably to the stellar continuum. We select a subsample with the highest surface brightness (SB) of H$\alpha$ as an indication of the dominance of the HII region of the recent starburst. This subsample should have the lowest contamination from old population and DIG. We also select a subsample with the lowest redshift. Given a fixed spatial resolution of the instrument, the lowest redshift indicates the best physical resolution and the least contamination from surrounding stellar populations. With these two subsamples, all trends of IMF variation discussed earlier still exist. Secondly, we compare our stellar metallicity with gas phase metallicity (i.e. oxygen abundance). Though our stellar metallicity is model dependent and digitalized, different gas phase metallicity indicators also have variation and inconsistency among them. Besides, stellar metallicity allows a more direct comparison with theoretical models while gas metallicity needs an arbitrary conversion to stellar metallicity before compared with models. Therefore, we choose to use stellar metallicity through out this work. We also test our results against three popular gas phase metallicity, namely O3N2 \citep{2004MNRAS.348L..59P}, N2O2 \citep{2002ApJS..142...35K}, and R23 \citep{2005ApJ...631..231P}. Our general results and conclusions do not change with gas phase metallicity.

In addition, the stochastic sampling of the IMF at actual star formation scale adds to the complexity of its nature. Massive star formation requires much denser molecular clouds than other stars, summarized as the "high mass star formation threshold" \citep[e.g.][]{2017MNRAS.466.3682B}. Then, the stochastic sampling of the IMF could potentially depend on local star formation rates, core mass function, etc. The concept of "integrated galactic initial mass function" has been proposed to bridge small scale physics of star formation with galactic stellar population \citep[e.g.][]{2003ApJ...598.1076K, 2017A&A...607A.126Y, 2018A&A...620A..39J}. More future work needs to be done in this direction.

Following the detailed discussion of potential systematics in \citet[][section 5.2]{2002A&A...394..443P}, we have considered several other effects, but none of them appears greatly relevant to our sample and analysis.

\subsection{Metallicity-dependent IMF variation}

There has been a rich history of studies, both observatoinal and 
theoretical, on the variation of stellar IMF. From the observatoinal 
side, a variety of probes have been used to constrain the IMF
in the past decade. 
\citep[e.g.][]{2003MNRAS.339L..12C,  2004MNRAS.355..728F, 
2010ApJ...709.1195T,2010Natur.468..940V,
Cappellari-12,2012ApJ...760...71C,2012MNRAS.422.2246M, 2012ApJ...753L..32S,2013ApJ...776L..26C,
2013MNRAS.429L..15F,2013MNRAS.433.3017L,2014ApJ...792L..37M,
2014ApJ...793...96S,2015MNRAS.446..493P,2016MNRAS.463.3220L,
Conroy-vanDokkum-Villaume-17,2017ApJ...838...77L,2017MNRAS.464..453D,
2017ApJ...841...68V,2017MNRAS.465..192Z,2018MNRAS.477.3954P,Watts-18,
2019MNRAS.489.5612D,2019MNRAS.485.5256Z}. 

Most studies have focused on early-type galaxies, but adopting
different techniques such as absorpition line spectroscopy
\citep[e.g.][]{2012ApJ...760...71C,Conroy-vanDokkum-Villaume-17},
kinematic analysis \citep[e.g.][]{Cappellari-12,2017ApJ...838...77L},
and stellar population synthesis
\citep[e.g.][]{2018MNRAS.477.3954P,2019MNRAS.485.5256Z}.  These
studies have well established that the high-mass end slope of  IMF is
positively correlated with the central stellar velocity dispersion
($\sigma_\ast$) of elliptical galaxies. Using integral field
spectroscopy of MaNGA, \citet{2017ApJ...838...77L} found that the
same correlation is well extended to bulges of  late-type
galaxies. More recently, also based on MaNGA data,
\citet{2018MNRAS.477.3954P} and \citet{2019MNRAS.485.5256Z}  adopted
different techniques and both found the IMF slope to be  also
positively correlated with stellar metallicity of the elliptical
galaxies.  \citet{2015ApJ...806L..31M} and \citet{2019MNRAS.485.5256Z} further demonstrated that,  when
compared to $\sigma_\ast$, stellar metallicity appears to be more
foundamental to the variation of the IMF slope. In an earlier work 
\citet{Zhang-07} found the same trend in star-forming galaxies, 
using an SDSS-based sample of WR spectra. Our work provides further 
observational evidence, particularly in the sense that the metallicity-dependent 
variation of the IMF slope holds even when binary population models 
are considered. 
% These studies echo the early discovery of very metal-poor 
% stars in the halo of our Galaxy \citep{2002Natur.419..904C,2005ARA&A..43..531B}, 
% which suggested the IMF is sensitive to even very low level of metal enrichment.
{ Besides, \citet{Zhang-18} and \citet{2019MNRAS.490.2838R} modeled isotopic ratios of CNO isotopes and discovered a top-heavy IMF is needed in their samples of starburst galaxies.}

In addition to these statistical studies of samples of galaxies,
detailed investigations of individual galaxies have also revealed
variation of IMF slope. For instance,
\citet{Conroy-vanDokkum-Villaume-17}  considered more flexible IMF
forms and obtained a super-Salpeter  IMF with an index of -2.7 for the
central region of NGC 1407,  a massive elliptical galaxy in their
sample. This result is  consistent with the expectation from the
$\alpha$-$Z$ correlation given the known mass-metallicity relation of
galaxies.  \citet{Watts-18} reported a top-light IMF with
$\alpha=-2.45$ for the dwarf irregular galaxy DDO 145 which has a low
metallicity of $Z=0.1\mathrm{Z_\odot}$. This result does not necessarily
conflict with  the positive correlation of $\alpha$ with $Z$ found
from samples of WR spectra, considering the large scatter of
individual WR galaxies/regions at fixed metallicity around the median
relation in \autoref{ew_z}.  

On the theoretical side, there have also been many studies on 
variations of the IMF \citep[e.g.][]{2006ApJ...642L..61T,
2008ApJ...672..757C,2008ApJ...682L...1M,
2009MNRAS.399.1255M,Hocuk-Spaans-2010a,Hocuk-Spaans-2010b,
2013MNRAS.435.2274W,Chabrier2014,2015MNRAS.448L..82F,2016MNRAS.462.2832C,
2018MNRAS.479.5448B,2018MNRAS.474.5259D,2018A&A...620A..39J,2019MNRAS.483..985B,
2019MNRAS.482.2515B,2019MNRAS.485.4852G,2019MNRAS.482..118G}. 
In particular, metallicity-dependent efficiency of molecular cloud
fragmentation has long been proposed to explain the IMF variations. 
In this case, when compared to its primordial counterpart, metal-enriched 
gas has more coolants which can keep the gas temperature lower and thus 
lead to more efficient fragementation, as shown in numerical simulations 
\cite[e.g.][]{2008ApJ...672..757C,2009MNRAS.399.1255M,Hocuk-Spaans-2010b}. 
For instance, \citet{Hocuk-Spaans-2010b} performed 38 
hydrodynamical simulations, exploring the metallicity dependence of molecular 
cloud fragmentation and possible variations in the dense core mass function. 
The results indicated a clear and strong dependence of fragmentation 
on metallicity, in the sense that the average fragment mass 
decreases with increasing metallicity. This is consistent with  
steeper (bottom-heavity) IMF slopes at higher metallicities, as 
observationally found in our work and previous studies. 

\subsection{ Importance of binary evolution for high mass stars or/and at low metallicities}
% \color{blue} 

%{\color{blue}
Observations show a positive correlation between binary fraction and stellar mass \citep[e.g.][]{2013ARA&A..51..269D, 2017ApJS..230...15M}.
Specifically for WR stars, \citet{vanderHucht-01} reported that 39\% of WR stars are found in binary systems.
This mass dependence consolidates the necessity of adopting binary evolution models in analyzing WR population.
Besides, dependence of binary fraction on metallicity exists as well. %}
\autoref{ew_z} shows that, at lowest metallicites ($Z<0.005$) 
the models of single stellar populations predict too weak a WR blue bump 
than the data, even when adopting a shallow IMF slope of $\alpha=2.00$ 
(which is the shallowest slope available in {\tt BPASS}). 
When considering the binary population models, the predicted 
WR features become more consistent with the data and the model of 
$\alpha=2$ well matches the median of the data points at the lowest
metallicities. This implies that binary stars prefer low-metallicity 
environment, although examinations of binary models of even 
shallower IMF's would be needed before one could make a solid conclusion. 
In fact, higher binary fractions have been both observationally found 
\citep[e.g.][]{2005ApJ...625..825L}, and theoretically predicted 
\citep[e.g.][]{2009MNRAS.399.1255M} for metal-poor stars. 
As pointed out by \citet{Guseva-Izotov-Thuan-00}, theoretical
models of \citet{Schaerer-Vacca-98} predicted that WR stars 
in massive close binary systems in the late phases of an instantaneous 
burst of star formation may have larger equivalent width. 
Therefore, a higher fraction of binary stars at low metallicites 
can more effectively increase the equivalent width of the WR bump,
thus bringing the model prediction to better matching the data.
This expectation also holds in {\tt BPASS} models, as 
can be seen in the right-hand panel of \autoref{ew_z}.  

\section{Conclusions}
\label{sec:con}
In summary, this work attempts to constrain the slope of the high-mass end of 
the stellar initial mass function (IMF) by comparing the spectroscopic 
features of Wolf-Rayet (WR) galaxies/regions with predictions of 
stellar population models that cover a range of IMF slopes and 
stellar metallicity. We combine three large spectroscopic 
samples: two based on SDSS sigle-fiber spectroscopy 
\citep{Zhang-07, Brinchmann-Kunth-Durret-08} and one based on the 
integral field spectroscopy from MaNGA \citep{liang_2020}. Our sample includes 
910 unique WR spectra. We measure the WR feature in each spectrum, 
as quantified by $EW_{4650}$ (equivalent width of the WR bump 
at around 4650\AA), and make comparisons with the single stellar 
population models of {\tt Starburst99} and {\tt BPASS}, as well as
the binary population models of {\tt BPASS}. We also obtain 
Bayesian evidence for different models by performing full spectral 
fitting to each spectrum using {\tt BIGS} \citep[][see \autoref{bigs} for a summary ]{Zhou2020a}.
The comparison between data and model predictions and the 
Bayesian evidence of different models allow us to effectively 
constrain the IMF slope $\alpha$ for WR galaxies/regions with 
different stellar metallicities.

Our conclusions can be summarized as follows. 
\begin{itemize}
    \item Comparisons of the observed $EW_{4650}$ with predictions 
    of single stellar population models from both {\tt Starburst99} 
    and {\tt BPASS} suggest a positive correlation of IMF slope 
    $\alpha$ with stellar metallicity $Z$, i.e. with steeper IMF 
    (more bottom-heavy) at higher metallicities. Specifically, an IMF with $\alpha$=1.00 is preferred at the lowest metallicity (Z$\sim$0.001), and an Salpeter or even steeper IMF is preferred at the highest metallicity (Z$\sim$0.03).
    \item Analysis of Bayesian evidence also implies steeper IMF's 
    at higher metallicities. 
    \item The above results hold when binary population models 
    are considered. In other words, any single IMF slope cannot 
    simultaneously explain the WR features of galaxies/regions 
    at all metallicities, in models of both singular populations 
    and binary populations. 
\end{itemize}

% With WR regions, we also have the potential to constrain the binary properties for the high-mass end.

\acknowledgments

We would like to thank the anonymous referee for the useful comments that have improved the paper.

This work is supported by the National Key R\&D Program of China
(grant No. 2018YFA0404502) and the National Science Foundation
of China (grant Nos. 11821303, 11973030, 11761131004, 11761141012, and 11603075).

Funding for SDSS-IV has been provided by the Alfred P. Sloan Foundation and Participating Institutions. Additional funding towards SDSS-IV has been provided by the US Department of Energy Office of Science. SDSS-IV acknowledges support and resources from the Centre for High-Performance Computing at the University of Utah. The SDSS web site is www.sdss.org.

SDSS-IV is managed by the Astrophysical Research Consortium for the Participating Institutions of the SDSS Collaboration including the Brazilian Participation Group, the Carnegie Institution for Science, Carnegie Mellon University, the Chilean Participation Group, the French Participation Group, Harvard–Smithsonian Center for Astrophysics, Instituto de Astrofsica de Canarias, The Johns Hopkins University, Kavli Institute for the Physics and Mathematics of the Universe (IPMU)/University of Tokyo, Lawrence Berkeley National Laboratory, Leibniz Institut fur Astrophysik Potsdam (AIP), Max-Planck-Institut fur Astronomie (MPIA Heidelberg), Max-Planck-Institut fur Astrophysik (MPA Garching), Max-Planck-Institut fur Extraterrestrische Physik (MPE), National Astronomical Observatory of China, New Mexico State University, New York University, University of Notre Dame, Observatario Nacional/MCTI, The Ohio State University, Pennsylvania State University, Shanghai Astronomical Observatory, United Kingdom Participation Group, Universidad Nacional Autonoma de Mexico, University of Arizona, University of Colorado Boulder, University of Oxford, University of Portsmouth, University of Utah, University of Virginia, University of Washington, University of Wisconsin, Vanderbilt University and Yale University.

\bibliography{wr_sec}
\bibliographystyle{aasjournal}

%\appendix
%\input{wr_icon}

\end{document}